\documentclass[twocolumn,showpacs,preprintnumbers,amsmath,amssymb]{revtex4}
\usepackage{graphicx}% Include figure files
\usepackage{color}% colored text for annotations/comments
\usepackage{dcolumn}% Align table columns on decimal point
\usepackage{bm}% bold math\newcommand{\TR}{\mbox{Tr}}
\usepackage{lipsum}

\newcommand{\op}[1]{\hat{#1}}
\renewcommand{\vec}[1]{\mathbf{#1}}

\newcommand{\energy}{E}
\newcommand{\tr}{}
\newcommand{\eqnref}[1]{Eq.~(\ref{#1})}
\newcommand{\EF}{{E_{\rm F}}}

\begin{document}

\preprint{APS/123-QED}

\title{Composition-dependent magnetic response properties of Mn$_{1-x}$Fe$_x$Ge alloys}% Force line breaks with \\

%\author{}
\author{S.~Mankovsky, S.~Wimmer, S.~Polesya, and H.~Ebert}
\affiliation{%
Dept. Chemie/Phys. Chemie, LMU Munich,
Butenandtstrasse 11, D-81377 Munich, Germany 
}%

\date{\today}% It is always \today, today,
             %  but any date may be explicitly specified

\begin{abstract}
    	The composition-dependent behavior of the Dzyaloshinskii-Moriya
	interaction (DMI), the spin-orbit torque (SOT), as well as anomalous
	and spin Hall conductivities of Mn$_{1-x}$Fe$_x$Ge alloys have been
	investigated by first-principles calculations using the relativistic
	multiple scattering Korringa-Kohn-Rostoker (KKR) formalism. The
	$D_{\rm xx}$ component of the DMI exhibits a strong dependence on the Fe
	concentration, changing sign at $x \approx 0.85$ in line with previous
	theoretical calculations as well as with experimental results
	demonstrating the change of spin helicity at $x \approx 0.8$.  A
	corresponding behavior with a sign change at $x \approx 0.5$ is
	predicted also for the Fermi sea contribution to the SOT, as this
	is closely related to the DMI.  In the case of anomalous and spin Hall
	effects it is shown that the calculated Fermi sea contributions are
	rather small and the composition-dependent behavior of these effects
	are determined mainly by the electronic states at the Fermi level. The
	spin-orbit-induced scattering mechanisms responsible for both these
	effects suggest a common origin of the minimum of the AHE and the sign
	change of the SHE conductivities.
\end{abstract}

% PACS, the Physics and Astronomy
                             % Classification Scheme.
%\keywords{Suggested keywords}%Use showkeys class option if keyword
                              %display desired
\maketitle

\section{ Introduction \label{IN}}

During the last decade Skyrmionic magnetic materials have moved into the focus
of scientific interest because their unique properties hold promises for
various applications in magnetic storage and spintronic devices~\cite{KST17}.
The key role for the formation of a Skyrmion magnetic texture is played by the
Dzyaloshinskii-Morirya interaction (DMI)~\cite{Dzy58, Mor60}. Its competition
with the isotropic exchange interaction, magnetic anisotropy, and the Zeeman
interaction in the presence of an external magnetic field determines the size
of Skyrmions and the region of stability in the corresponding phase diagram.
Another important characteristic feature of Skyrmions is their helicity (i.e.,
the spin spiraling direction), which is determined by the orientation of the
involved Dzyaloshinskii-Morirya interaction vectors and can be exploited as an
additional degree of freedom for the manipulation of
Skyrmions~\cite{DT16,TOY+01,BDT04}. The correlation between the Skyrmion
helicity and crystal chirality has been already discussed in the
literature~\cite{IEM+85,DGM+11}. Recent experiments have demonstrated in
addition a change of the Skyrmion helicity with the chemical composition in
the case of B20 alloys~\cite{SYH+13,GSA+14} while the crystal chirality was
unaltered. This finding opens an alternative possibility for DMI engineering in
order to manipulate the Skyrmion helicity.

This holds particularly true for the Mn$_{1-x}$Fe$_x$Ge alloy system which
is in the center of interest for the present investigation.  Experimentally, it
was found~\cite{SYH+13,GPS+13} that the size of Skyrmions in this material can
be tuned by changing the Fe concentration, reaching a maximum at
$x\sim0.8$~\cite{GPS+13}, i.e., at the concentration when the Skyrmion helicity
changes sign without a change of the crystal chirality. This behavior was
investigated theoretically~\cite{GFS+15, KNA15} via first-principles
calculations of the DMI and analyzing the details of the electronic structure
that may have an influence on it. Gayles \emph{et al.}~\cite{GFS+15} have
demonstrated that the sign of the DMI in Mn$_{1-x}$Fe$_x$Ge can be explained by
the relative positions in energy of the $d^\uparrow_{\rm xy}$- and
$d^\downarrow_{x^2-y^2}$-states of Fe which change when the Fe concentration
increases above $x \sim 0.8$. As a consequence a flip of the chirality of the
magnetic texture occurs. Similar conclusions have been drawn by Koretsune et
al. \cite{KNA15}.  While these calculations have been done treating chemical
disorder within the virtual crystal approximation (Ref.~\onlinecite{GFS+15}) or
even employing the rigid band approximation (Ref.~\onlinecite{KNA15}), the
present work is based on the coherent potential approximation (CPA) alloy
theory, which should give more reliable results for the electronic structure of
disordered alloys. 

In addition we investigate the concentration dependence of response properties
connected to spin-orbit coupling (SOC) in the presence of an applied electric
field, i.e., the spin-orbit torque (SOT), the anomalous Hall effect (AHE) and
the spin Hall effect (SHE),  as these are important for practical applications.
Especially, we focus on the SOT, expecting common features with the DMI
according to recent findings by Freimuth et al. \cite{FBM14}.

%Manipulation of the spin helix in FeGe thin films and FeGe/Fe
%multilayers \cite{PST+15}
%Extended Skyrmion Phase in Epitaxial FeGe(111) Thin Films \cite{HC12}

The article is organised as follows. We start with theoretical details on the
formalisms employed to calculate DMI parameters and linear response
coefficients from first principles in section~\ref{TB}. Results for the
Mn$_{1-x}$Fe$_x$Ge alloy system are presented and discussed in
section~\ref{RD}, subdivided into Dzyaloshinskii-Morirya interaction
(\ref{ssec:DMI}), spin-orbit torque (\ref{ssec:SOT}), anomalous and spin Hall
conductivity (\ref{ssec:Hall}), and symmetry considerations (\ref{ssec:Sym}).
We conclude with a brief summary in (\ref{Sum}). Additional derivations
connected to the expressions in \ref{TB} are given in the Appendix~\ref{App}.

\section{Theoretical details \label{TB}}

All calculations were performed using the fully relativistic
Korringa--Kohn--Rostoker (KKR) Green function method \cite{SPRKKR7,EKM11}
within the framework of local spin density approximation (LSDA) to density
functional theory (DFT) and the parametrization scheme for the exchange and
correlation potential as given by Vosko et al. \cite{VWN80}. A cutoff $l_{max}
= 3$ was used for the angular momentum expansion of the Green function. The
chemical disorder was treated within the  coherent potential approximation
(CPA) alloy theory \cite{Sov67,Vel69}.

In order to investigate the composition dependent behavior of
the Skyrmion size and helicity observed in experiment, we have calculated 
the $D_{\rm xx}$ element of the micromagnetic DMI tensor as 
a function of Fe concentration $x$. 
As it was demonstrated previously \cite{ME17}, this quantity can be
calculated in two different ways. Either by performing a direct evaluation of
the expression
 
%WWWWWWWWWWWWWWWWWWWWWWWWWWWWWWWWWWWWWWWWWWWWWWWWWWWWWWWWWWWWWWWWWWWWWWWWWw
\begin{widetext}
 %EEEEEEEEEEEEEEEEEEEEEEEEEEEEEEEEEEEEEEEEEEEEEEEEEEEEEEEEEEEEEEEEEEEEEEEEEE
 \begin{eqnarray}
  D_{\mu\nu} &=& \frac{1}{\pi} \mbox{Re}\, \mbox{Tr}\, \int^\EF dE \,(E - \EF)
	 \nonumber \\
	 &&\times  \frac {1}{\Omega_{BZ}} \int d^3k \, \Big[ \underline{O}(E)\,
	 \underline{\tau}(\vec{k},E) \,
	 \underline{T}_{\mu}(E)\,\frac{\partial}{\partial k_\nu}\,
	 \underline{\tau}(\vec{k},E) -  \underline{T}_{\mu}(E) \,
	 \underline{\tau}(\vec{k},E) \,
 \underline{O}(E)\,\frac{\partial}{\partial k_\nu}\, \underline{\tau}(\vec{k},E)
 \Big] \;.
 \label{Eq:Dxx_tau} 
 \end{eqnarray}
 %EEEEEEEEEEEEEEEEEEEEEEEEEEEEEEEEEEEEEEEEEEEEEEEEEEEEEEEEEEEEEEEEEEEEEEEEEE
 %
 with the overlap integrals and the matrix elements of the torque operator
 $\op{T}_{\mu} = \beta[\boldsymbol{\sigma} \times \hat{z}]_{\mu} B_{xc}(\vec{r})$
 \cite{EM09a}
 \begin{eqnarray}
	 \left[ \underline{O} \right]_{\Lambda\Lambda'} &=& \int_{\Omega} d^3 r Z^\times_\Lambda (\vec{r},E) Z^{j}_{\Lambda'} (\vec{r},E) \nonumber \\
 \left[ \underline{T}_{\mu} \right]_{\Lambda\Lambda'} &=& \int_{\Omega} d^3 r Z^\times_\Lambda (\vec{r},E) \hat{T}_\mu Z^{j}_{\Lambda'} (\vec{r},E) \;,
 \label{Eq:OnTLaLap} 
 \end{eqnarray}
 or by using the interatomic $D_\mu^{ij}$ interactions
 %
 %EEEEEEEEEEEEEEEEEEEEEEEEEEEEEEEEEEEEEEEEEEEEEEEEEEEEEEEEEEEEEEEEEEEEEEEEEE
 \begin{eqnarray}
  D_{\mu\nu}  &=&  \sum_{ij} D_\mu^{ij} \, (\vec{R}_j - \vec{R}_i)_\nu
 \label{Eq:DMI_micmag_vs_Heis}
 \end{eqnarray}
 %EEEEEEEEEEEEEEEEEEEEEEEEEEEEEEEEEEEEEEEEEEEEEEEEEEEEEEEEEEEEEEEEEEEEEEEEEE
%
 that are calculated in an analogous way \cite{ME17}.

 The current-induced torkance \cite{WCS+16} and the anomalous \cite{LKE10} and
 spin \cite{LGK+11} Hall conductivities were calculated within the Kubo linear
 response formalism using the expression
%EEEEEEEEEEEEEEEEEEEEEEEEEEEEEEEEEEEEEEEEEEEEEEEEEEEEEEEEEEEEEEEEEEEEEEEEEE
 \begin{eqnarray}
    \label{Eq:Bastin-O1-O2}
    {\cal R}_{\mu\nu}  \nonumber     &=&
    {{\cal R}_{\mu\nu}^{I}}+{{\cal R}_{\mu\nu}^{II}} \nonumber    \\
    &=&
      { - \frac{\hbar }{4\pi\Omega}\int_{-\infty}^{\infty}
    \frac{df(\energy)}{d\energy}
    \mbox{Trace}
    \left<
      \op{B}_{\mu}
      (\op{G}^{+}-\op{G}^{-})
      \op{A}_{\nu}
      \op{G}^{-}
      -\op{B}_{\mu}
      \op{G}^{+}
      \op{A}_{\nu}
      (\op{G}^{+}-\op{G}^{-})
    \right>
    {\rm d}
    \energy
    }
  \nonumber  \\
    % ---------------------------------------------------------------------
    &&
   {
    \label{Eq:Bastin-O1-O2-2}
    + \frac{\hbar }{4\pi\Omega}\int_{-\infty}^{\infty}
    f(\energy)
    \mbox{Trace}
    \left<
      {\bigg(}
      \op{B}_{\mu} \op{G}^{+}\op{A}_{\nu}
      \frac{d\op{G}^{+}}{d\energy}
      -\op{B}_{\mu}\frac{d\op{G}^{+}}{d\energy}  \op{A}_{\nu} \op{G}^{+}
        {\bigg)}
          \,
      -
       \,
      {\bigg(}
      [G^-] 
      {\bigg)}
    \right>
    {\rm d}
    \energy
} \;,
  \end{eqnarray}
%EEEEEEEEEEEEEEEEEEEEEEEEEEEEEEEEEEEEEEEEEEEEEEEEEEEEEEEEEEEEEEEEEEEEEEEEEE
where ${\cal R}_{\mu\nu}^{I}$ and ${\cal R}_{\mu\nu}^{II}$ are Fermi surface
and Fermi sea contributions, respectively.  The operator $\op{A}_{\nu}$
representing in all three cases the perturbation is the electric current
density operator $\op{j}_{\nu} = -|e|c\alpha_{\nu}$. For the calculations of
the anomalous Hall conductivity one has for the response $\op{\bf B} = \op{\bf
A}$, for the spin Hall conductivity $\op{\bf B} = \op{\bf P}\op{\bf A}$ with
the relativistic spin-polarization operator $\op{\bf P}$ \cite{BW48,VGW07},
while for the calculations of the spin-orbit torkances $t_{\rm \mu\nu}$ the
torque operator $\op{B}_{\mu} = \op{T}_{\mu}$ has to be used.  Additional
calculations for the Fermi sea torkance have been performed following the
relationship between this quantity and the DMI parameters as suggested by
Freimuth et al.  \cite{FBM14}. In line with \eqnref{Eq:Dxx_tau}, these
calculations were based on the expression
%EEEEEEEEEEEEEEEEEEEEEEEEEEEEEEEEEEEEEEEEEEEEEEEEEEEEEEEEEEEEEEEEEEEEEEEEEE
\begin{eqnarray}
 t^{\rm sea}_{\mu\nu}  &=&   -\frac{e}{\pi}\,\, \mbox{Re}\, \mbox{Tr}\, \int^\EF dE \, \nonumber \\ 
 &&\times  \frac {1}{\Omega_{BZ}} \int d^3k \, \Big[ \underline{O}(E)\,
\underline{\tau}(\vec{k},E) \,
\underline{T}_{\mu}(E)\,\frac{\partial}{\partial k_\nu}\,
\underline{\tau}(\vec{k},E) -  \underline{T}_{\mu}(E) \,
\underline{\tau}(\vec{k},E) \,
\underline{O}(E)\,\frac{\partial}{\partial k_\nu}\,
\underline{\tau}(\vec{k},E)  \Big] \;,
\label{Eq:txx_tau} 
\end{eqnarray}
%EEEEEEEEEEEEEEEEEEEEEEEEEEEEEEEEEEEEEEEEEEEEEEEEEEEEEEEEEEEEEEEEEEEEEEEEEE
\tr{which obviously differs, apart from prefactors, from \eqnref{Eq:Dxx_tau}
only by the weighting factor $(E - E_{\rm F})$.}  Both expressions for
$t^{\rm sea}_{\mu\nu}$ should be equivalent, as can be demonstrated for the
particular case of a translationally invariant system.  In this case the
relationship between \eqnref{Eq:txx_tau} and the Fermi sea term
$t^{\rm sea}_{\mu\nu}$ in \eqnref{Eq:Bastin-O1-O2} can be established using the
expression for the group velocity suggested by Shilkova and Shirokovskii
discussed below \cite{SST86,SS88,GFP+11} (see \ref{App}).
%EEEEEEEEEEEEEEEEEEEEEEEEEEEEEEEEEEEEEEEEEEEEEEEEEEEEEEEEEEEEEEEEEEEEEEEEEE

Alternatively, we have
%EEEEEEEEEEEEEEEEEEEEEEEEEEEEEEEEEEEEEEEEEEEEEEEEEEEEEEEEEEEEEEEEEEEEEEEEEE
\begin{eqnarray}
 t_{\mu\nu}  &=&  \sum_{ij} t_\mu^{ij} \, (\vec{R}_j - \vec{R}_i)_\nu    \;,
\label{Eq:SOT_micmag_vs_Heis}
\end{eqnarray}
%EEEEEEEEEEEEEEEEEEEEEEEEEEEEEEEEEEEEEEEEEEEEEEEEEEEEEEEEEEEEEEEEEEEEEEEEEE
with the interatomic torkance terms 
%EEEEEEEEEEEEEEEEEEEEEEEEEEEEEEEEEEEEEEEEEEEEEEEEEEEEEEEEEEEEEEEEEEEEEEEEEE
\begin{eqnarray}
  t_\mu^{ij} & = &
-\left(\frac{e}{2\pi} \right) \mbox{Im}\,\mbox{Tr} \int^\EF dE \,
%%% \nonumber \\     &\times & 
\sum_{\Lambda_1\Lambda_2\Lambda_3\Lambda_4} 
       \bigg[ { O^{j}_{\Lambda_4\Lambda_1}(E)} 
	       \,   {\tau}^{j i}_{\Lambda_1\Lambda_2}(E)  \, {T^{i}_{\mu,\Lambda_2
          \Lambda_3}(E)} \, {\tau}^{i j}_{\Lambda_3\Lambda_4}(E) \nonumber \\
  &&  \hspace{6cm} -  { O^{i}_{\Lambda_4 \Lambda_1}(E)} \,  {\tau}^{i
      j}_{\Lambda_1\Lambda_2}(E) \, {T^{j}_{\mu,\Lambda_2\Lambda_3}(E)}  \,
        {\tau}^{ji}_{\Lambda_3\Lambda_4}(E)\bigg] \;
        \label{Eq:DMI_tij}
\end{eqnarray}
%EEEEEEEEEEEEEEEEEEEEEEEEEEEEEEEEEEEEEEEEEEEEEEEEEEEEEEEEEEEEEEEEEEEEEEEEEE
\end{widetext}
%WWWWWWWWWWWWWWWWWWWWWWWWWWWWWWWWWWWWWWWWWWWWWWWWWWWWWWWWWWWWWWWWWWWWWWWWWW
that are obtained in analogy to the interatomic DMI parameters.

\section{Results and discussion \label{RD}}

\subsection{Dzyaloshinskii-Moriya interaction \label{ssec:DMI}}

\begin{figure}[htb]
\includegraphics[width=0.45\textwidth,angle=0,clip]{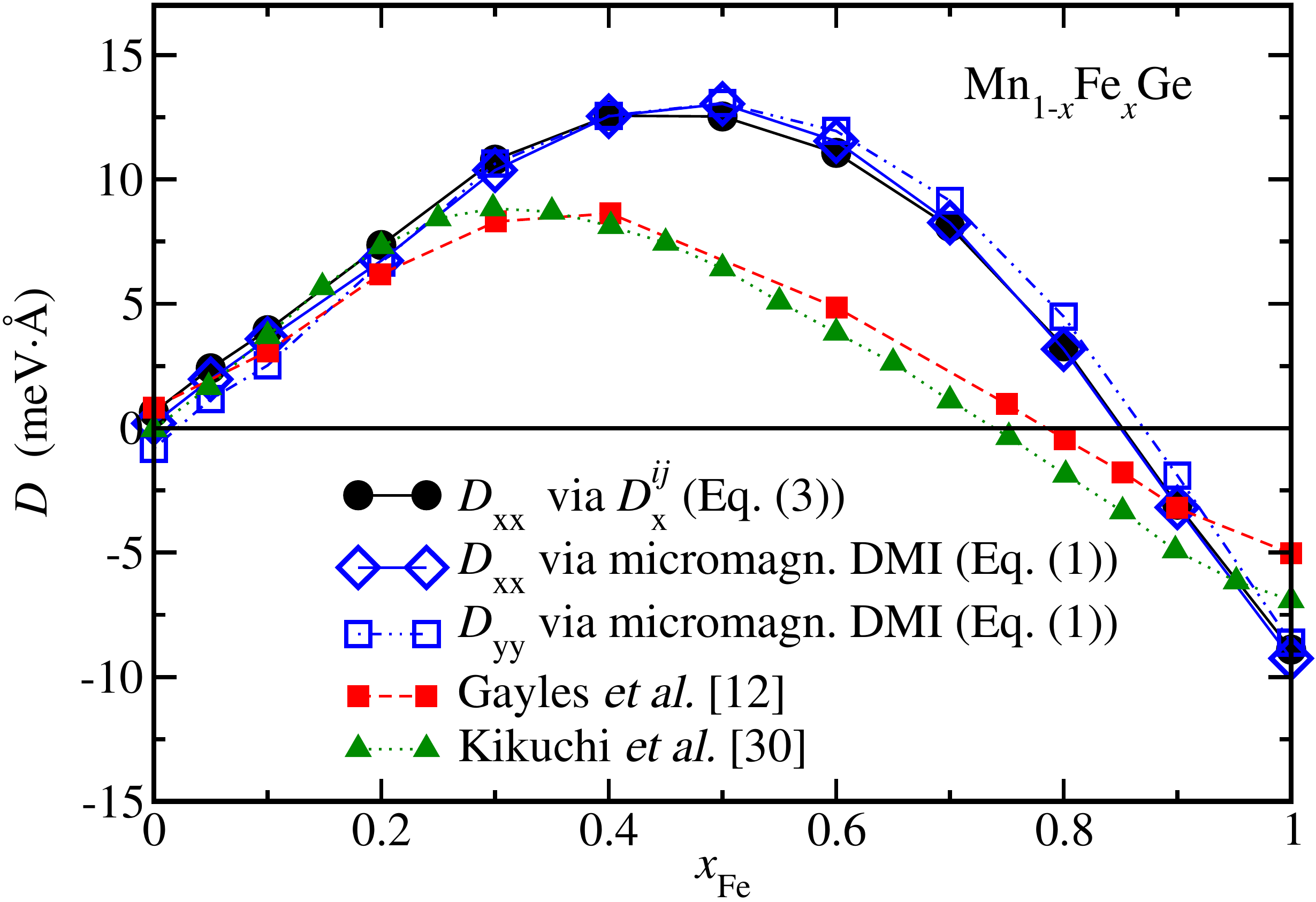}\;(a)
\includegraphics[width=0.45\textwidth,angle=0,clip]{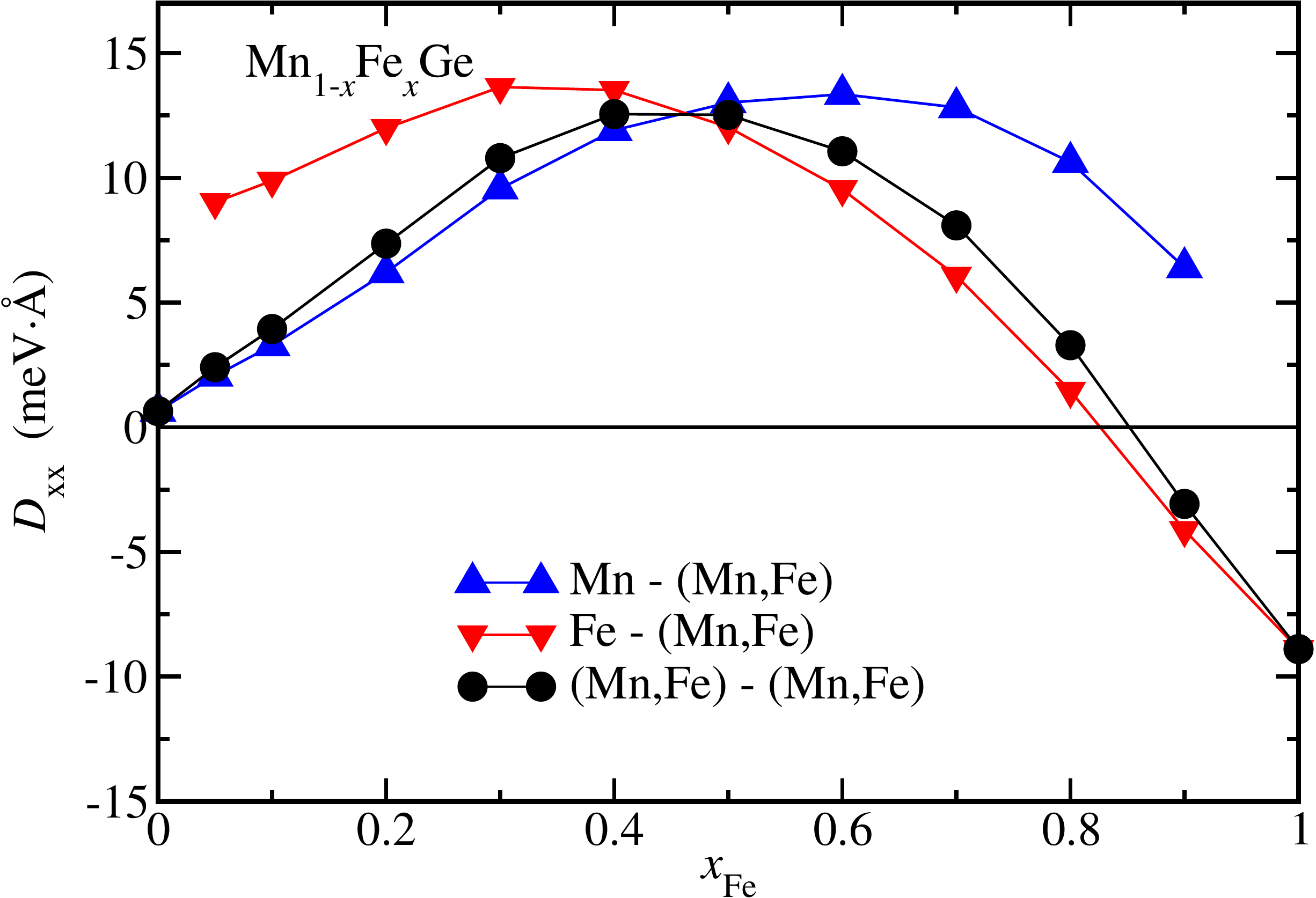}\;(b)
\caption{\label{fig:CMP_DMI} (a) Results for ${D}_{\rm xx}(x)$ in
	Mn$_{1-x}$Fe$_x$Ge calculated using Eq.\ (\ref{Eq:DMI_micmag_vs_Heis})
	(circles) and for ${D}_{\rm xx}(x)$ (diamonds) and
	${D}_{\rm yy}(x)$ (empty squares) calculated using using Eq.\ (\ref
	{Eq:Dxx_tau}) in comparison with the results of other calculations
	\cite{GFS+15} (filled squares) and \cite{KKAT16} (triangles). 
        (b) The element-resolved Dzyaloshinskii-Moriya interaction in
	Mn$_{1-x}$Fe$_x$Ge ${D}_{\rm xx}^{Mn}$ (triangles up) and ${D}_{\rm xx}^{Fe}$
	(triangles down). The total ${D}_{\rm xx}(x)$ function is again shown as
circles as in (a).}  
\end{figure}

In the following we first focus on the behavior of the DMI in
Mn$_{1-x}$Fe$_x$Ge as a function of Fe concentration $x$. The dependence of the
DMI parameter $D_{\rm xx}(x)$ on $x$ is plotted  in Fig.~\ref{fig:CMP_DMI}\,(a) in
comparison with available theoretical results from other groups
\cite{GFS+15,KKAT16}.  The results calculated using
%Eq.\ (\ref{Eq:DMI_micmag_vs_Heis})
an explicit expression for $D_{\rm xx}$ derived recently \cite{ME17} are given by
open diamonds, while those based on the interatomic interaction parameters
$\vec{D}^{ij}$  are given by solid circles.  Although the latter value has
contributions only from the $\vec{D}_{Fe-Fe}^{ij}$, $\vec{D}_{Mn-Mn}^{ij}$ and
$\vec{D}_{Fe-Mn}^{ij}$ interatomic DMI pair interaction terms, both results are
in very good agreement with each other. They also fit reasonably well to the
theoretical results by other groups shown by dashed \cite{GFS+15} and
dashed-dotted \cite{KKAT16} lines. The deviations between these and the present
work are most likely caused by the different approach used to treat the
chemical disorder in the alloy. As it was mentioned above, the CPA alloy theory
was used in the present work, while the previous results \cite{GFS+15,KKAT16}
have been obtained using the so-called virtual crystal approximation.  As it
follows from Fig.~\ref{fig:CMP_DMI}\,(a), $D_{\rm xx}(x)$ changes sign at $x
\approx 0.8$, in line with the experimental observation \cite{GPS+13}. 
%{\color{red} 
A very similar concentration dependence is also observed for the $D_{\rm yy}(x)$
component (open squares). The deviation from $D_{\rm xx}(x)$, that is allowed
by crystal symmetry (see subsection~\ref{ssec:Sym}), is itself a function of $x$
but small throughout.
%}
From the
element-projected plots shown in Fig.~\ref{fig:CMP_DMI}(b) one can see
that $D^{Fe}_{\rm xx}(x)$ and $D^{Mn}_{\rm xx}(x)$ have their maxima at a different Fe
concentration, i.e., at $x \approx 0.3$ and $x \approx 0.6$ for Fe and Mn,
respectively.  As one notes, ${D}_{\rm xx}^{Fe}(x)$ changes its sign at $x \sim
0.8$ if $x$ increases, while ${D}_{\rm xx}^{Mn}(x)$ does not change sign.

% %%%%%%%%%%%%%%%%%%%%%%%%%%%%%%%%%%%%%%%%%%%%%%%%%%%%%%%%%%%%%%%%%%%%%%%%%%%%%%%
% %%%%%%%%%%%%%%%%%%%%%%%%%%%%%%%%%%%%%%%%%%%%%%%%%%%%%%%%%%%%%%%%%%%%%%%%%%%%%%%

\begin{figure*}[htb]
\includegraphics[width=0.45\textwidth,angle=0,clip]{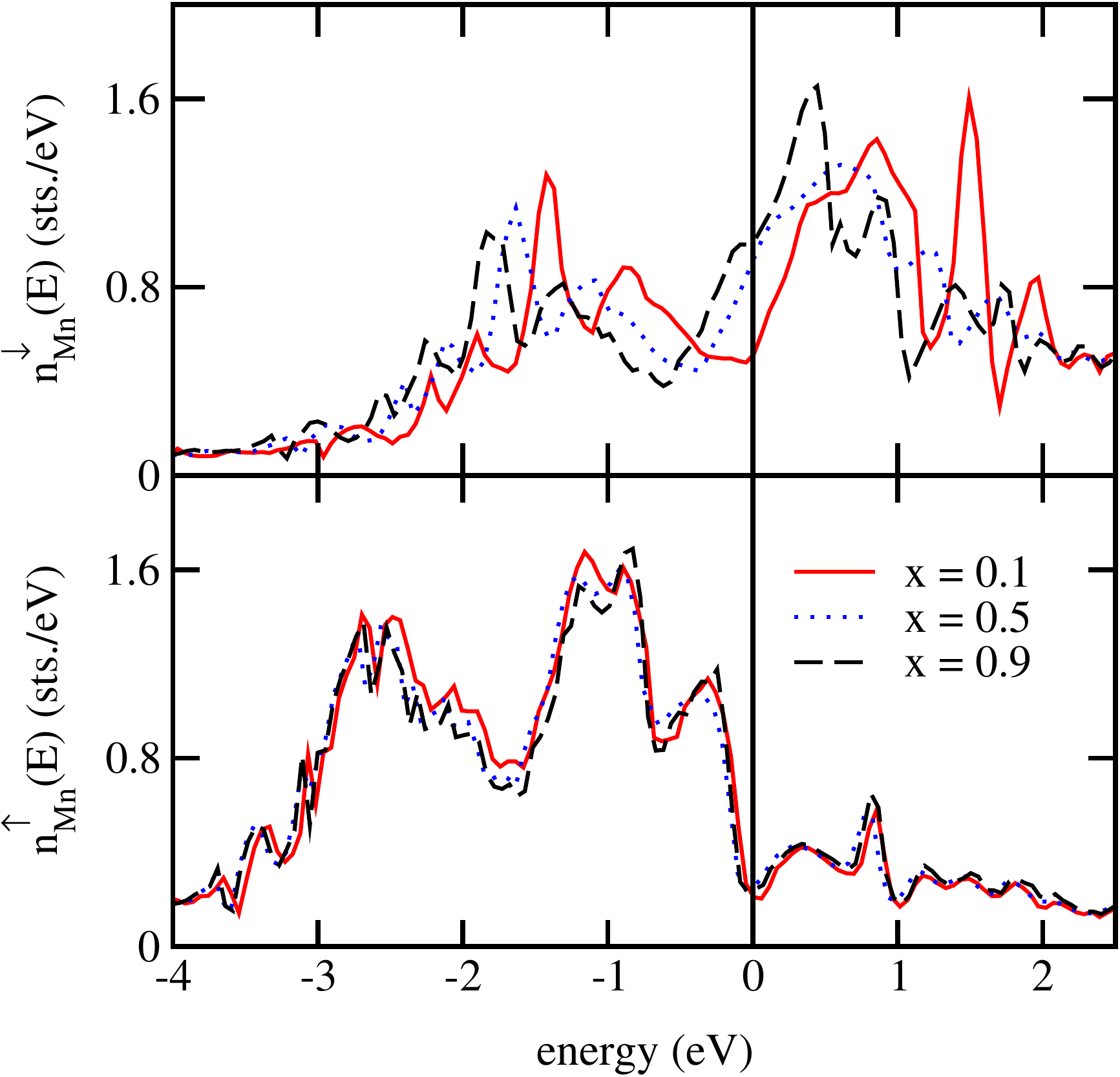}\;(a)
\includegraphics[width=0.45\textwidth,angle=0,clip]{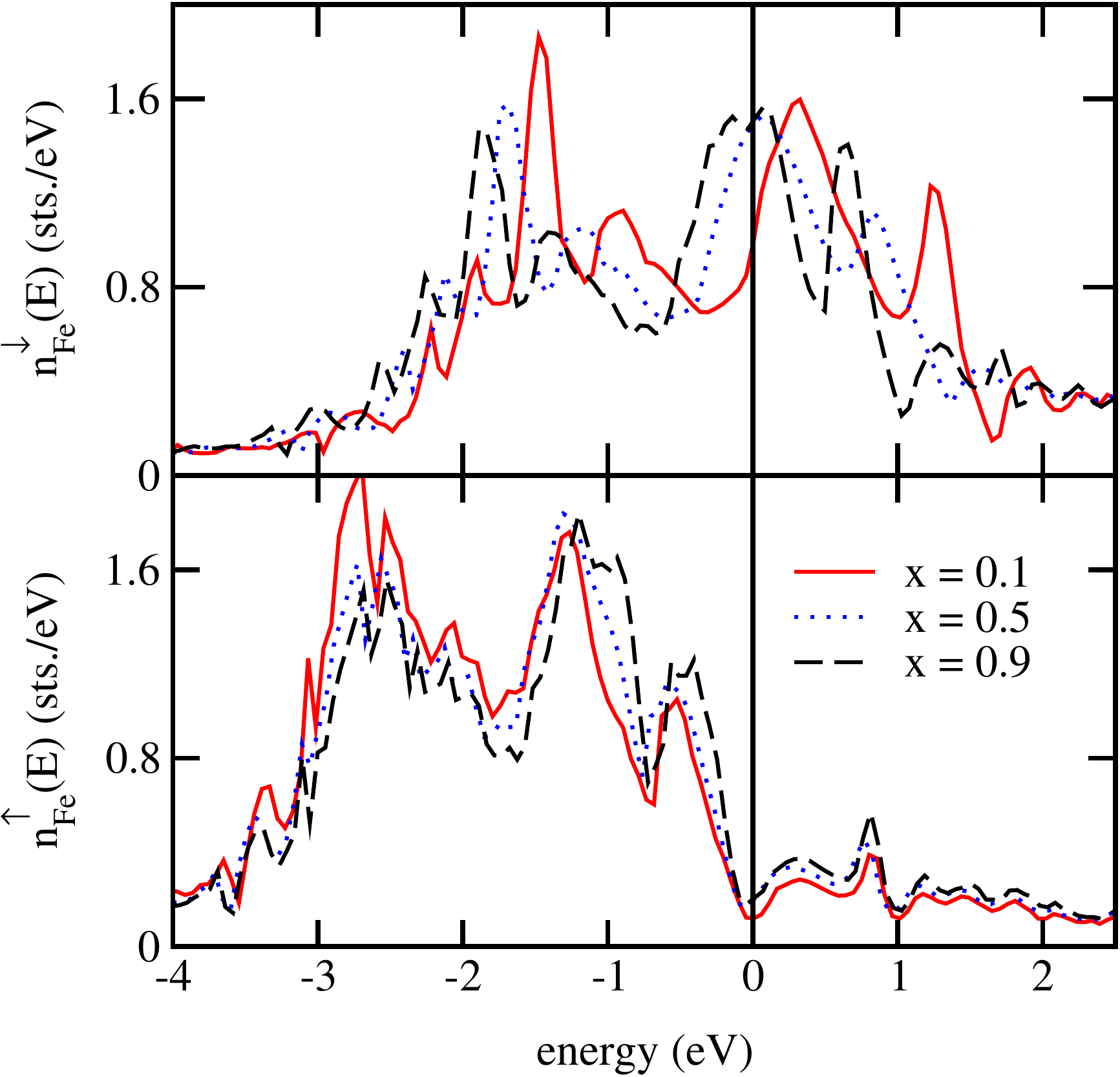}\;(b)
\caption{\label{fig:DOS_FeMnGe}  The spin- and element-resolved DOS on Mn (a)
and Fe (b) atoms in Mn$_{1-x}$Fe$_x$Ge for $x = 0.1, 0.5$ and $0.9$.}   
\end{figure*}

%{\color{red} 
The observed concentration dependence of the DMI was associated in the
literature \cite{GFS+15, KNA15,KKAT16} with specific features of the electronic
structure and their modification with the Fe concentration $x$.
Fig.~\ref{fig:DOS_FeMnGe} shows corresponding results of electronic structure
calculations making use of the CPA alloy theory, i.e., the spin- and
element-resolved density of states (DOS) on Mn (a) and Fe (b) sites in
Mn$_{1-x}$Fe$_x$Ge for the three different concentrations $x = 0.1, 0.5$, and
$0.9$.
%
%
% %%%%%%%%%%%%%%%%%%%%%%%%%%%%%%%%%%%%%%%%%%%%%%%%%%%%%%%%%%%%%%%%%%%%%%%%%%%%%%%
% %%%%%%%%%%%%%%%%%%%%%%%%%%%%%%%%%%%%%%%%%%%%%%%%%%%%%%%%%%%%%%%%%%%%%%%%%%%%%%%%
% %%%%%%%%%%%%%%%%%%%%%%%%%%%%%%%%%%%%%%%%%%%%%%%%%%%%%%%%%%%%%%%%%%%%%%%%%%%%%%%
\begin{figure*}[]
 \includegraphics[width=0.45\textwidth,angle=0,clip]{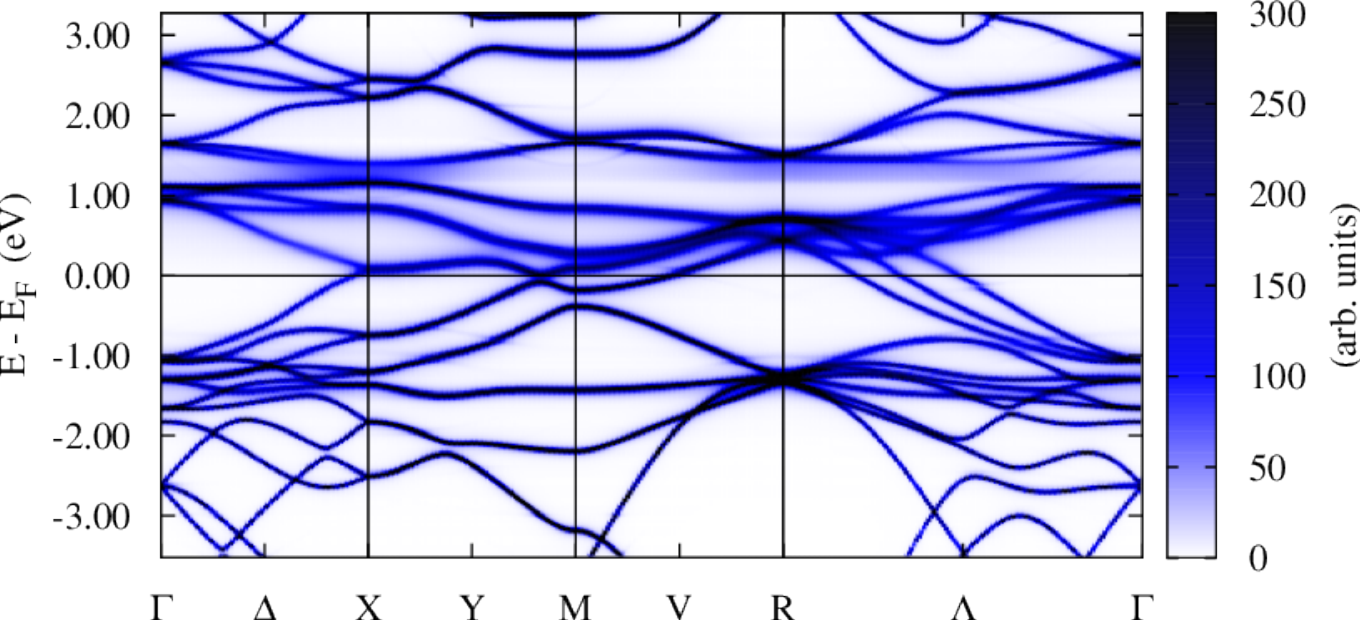}\;(a)
 \includegraphics[width=0.45\textwidth,angle=0,clip]{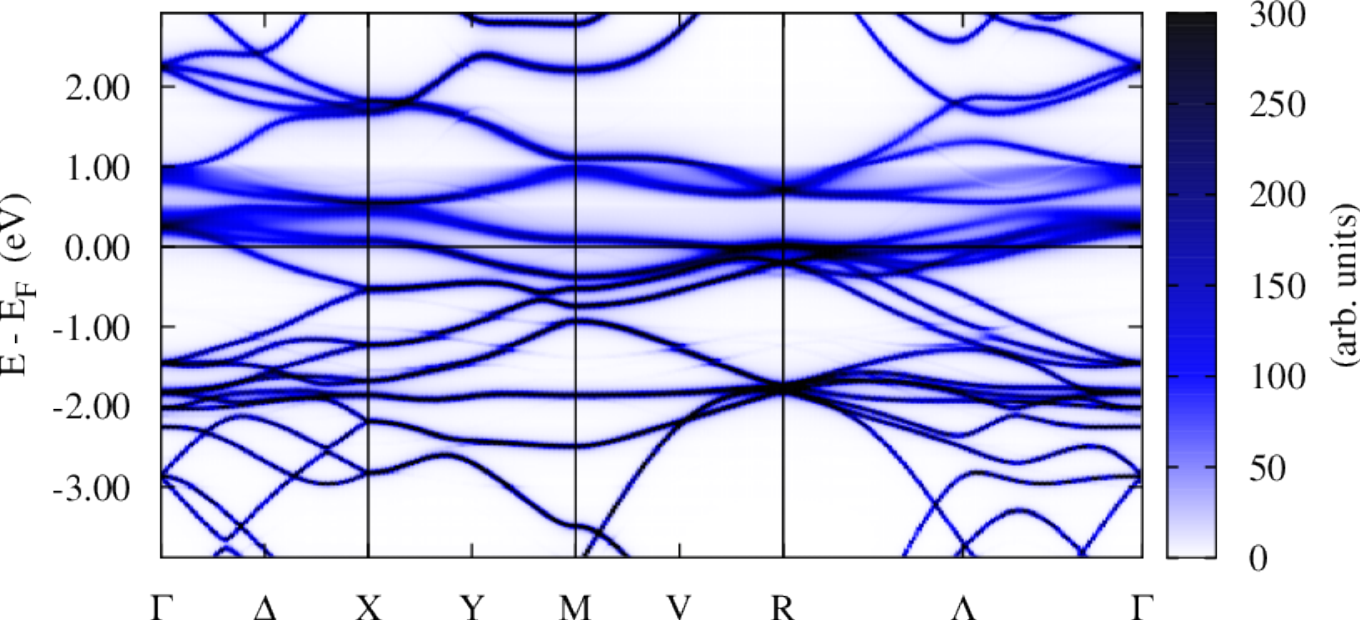}\;(b)
 \includegraphics[width=0.45\textwidth,angle=0,clip]{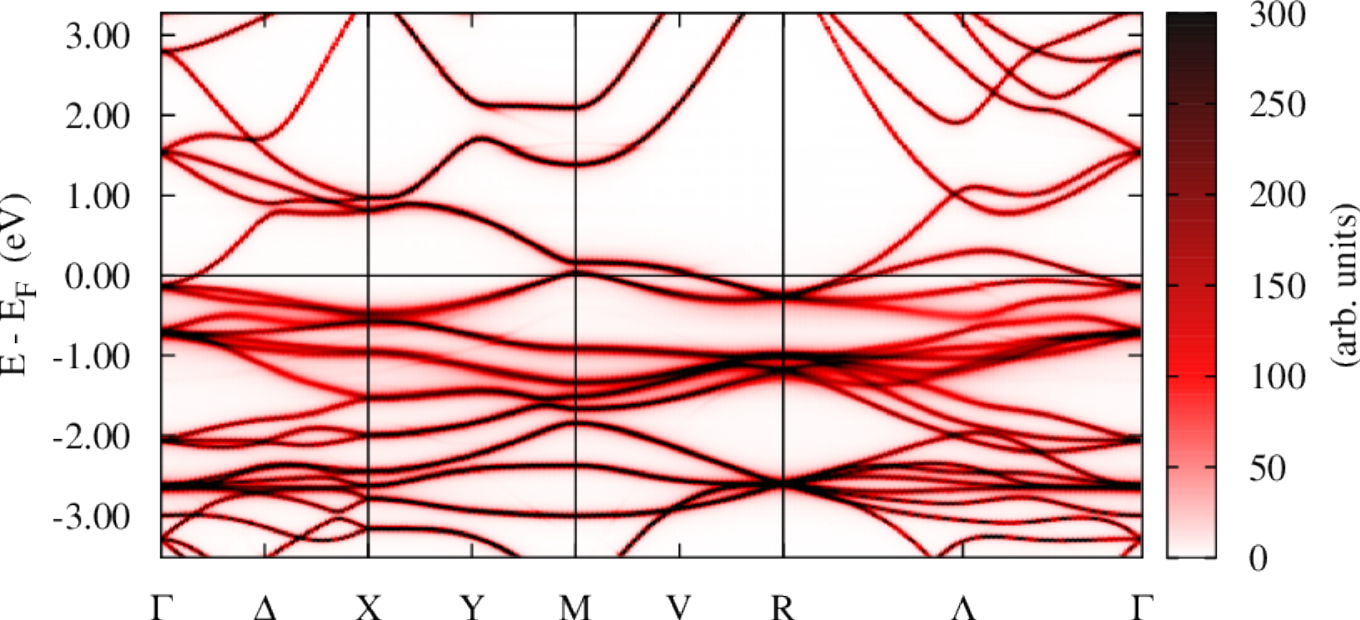}\;(c)
 \includegraphics[width=0.45\textwidth,angle=0,clip]{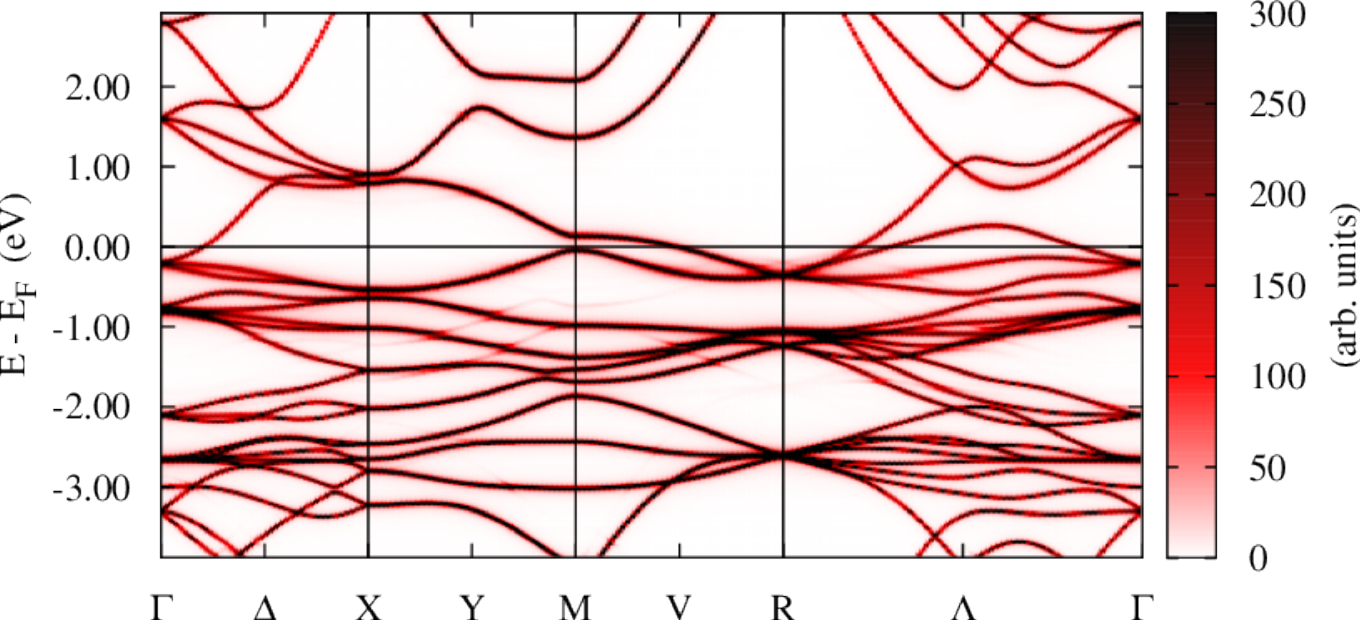}\;(d)
\caption{\label{fig:BSF_FeMnGe} The spin-resolved Bloch spectral function in
Mn$_{0.9}$Fe$_{0.1}$Ge and Mn$_{0.1}$Fe$_{0.9}$Ge for minority- (a) and
(b) and majority-spin (c) and (d) states, respectively.}   
\end{figure*}
% %%%%%%%%%%%%%%%%%%%%%%%%%%%%%%%%%%%%%%%%%%%%%%%%%%%%%%%%%%%%%%%%%%%%%%%%%%%%%%%
%
As one can see in the bottom panels, the occupied majority-spin states of Mn
and Fe are very close to each other and hardly depend on the Fe concentration.
Obviously, chemical disorder has only a weak impact for this spin subsystem,
leading to a rather weak disorder-induced smearing of the energy bands. This
can be seen as well in Fig.~\ref{fig:BSF_FeMnGe}\,(c) and (d), that show the
Bloch spectral function for majority-spin states in Mn$_{0.9}$Fe$_{0.1}$Ge and
Mn$_{0.1}$Fe$_{0.9}$Ge, respectively. On the other hand, the different exchange
splitting for the electronic states on Mn and Fe sites leads to different
positions for their minority-spin states and as a consequence to a pronounced
disorder-induced smearing of the energy bands for the disordered
Mn$_{1-x}$Fe$_x$Ge alloys.  Again this can be seen in the upper panels of
Fig.~\ref{fig:DOS_FeMnGe}\,(a) and (b), as well as in
Fig.~\ref{fig:BSF_FeMnGe}\,(a) and (b), showing the Bloch spectral function for
minority-spin states in case of $x = 0.1$ and $0.9$, respectively.  Moreover,
the exchange splitting for Fe and Mn both decreases upon an increase of the Fe
concentration.  
%This can
%clearly be seen in Fig.~\ref{fig:DOS_FeMnGe} which displays the
%element-projected DOS for Mn and Fe atoms individually. 
As a consequence, the Fe and Mn spin magnetic moments decrease simultaneously
as can be seen in Fig.~\ref{fig:M_spin}. 
%}

Fig.~\ref{fig:DOS_FeMnGe} indicates that the concentration-dependent
modification of the electronic structure has two-fold character. First, the Fe
minority-spin $d^{\downarrow}_{x^2-y^2}$ states move down in energy from their
position above the Fermi level at small Fe concentration ($x = 0.1$, solid line
in Fig.~\ref{fig:DOS_FeMnGe}(b)) to a position below the Fermi energy at high
Fe content ($x = 0.9$, dashed line in Fig.~\ref{fig:DOS_FeMnGe}(b)).
Additionally, a weak shift of the majority-spin $d^{\uparrow}_{\rm xy}$-states
of Fe towards the Fermi energy can be observed.  This behavior, as discussed
previously \cite{KNA15, GFS+15}, leads to a sign change of the Fe-projected as
well as the total DMI at $x \sim 0.8$.  At the same time
Fig.~\ref{fig:DOS_FeMnGe}\,(a) shows that the minority-spin
$d^{\downarrow}_{x^2-y^2}$-states of Mn stay essentially unoccupied over the
whole concentration range. As a consequence, ${D}_{\rm xx}^{Mn}(x)$ does not
exhibit any sign changes.  \if{}Secondly, the composition of Mn$_{1-x}$Fe$_x$Ge
has an impact on the electronic structure which is related to the change of the
relative contributions of the Fe and Mn states varying with concentration from
leading contribution of the Mn states (Mn rich limit) to one of the Fe states
(Fe rich limit).\fi As the positions of the element-projected minority-spin
states of Fe and Mn are rather different (see Fig.~\ref{fig:DOS_FeMnGe}), the
increase of the contributions of minority-spin Fe states with increasing $x$ in
parallel with the decreasing contribution of corresponding Mn states leads for
the alloy system to an apparent shift of the electronic energy bands. According
to Refs.~\onlinecite{GFS+15} and \onlinecite{KNA15} this should also lead to a
sign change of the DMI parameter.

Finally, it is worth to mention that there are different trends in the
behavior of the DMI parameter in the Mn-rich limit when comparing theoretical
results (both, present and previous) with experimental data~\cite{GPS+13}.
As it was remarked by Gayles \emph{et al.} \cite{GFS+15} the origin of this
difference is not clear and the authors suggest certain mechanisms to be
responsible for that. We would like to add here that the micromagnetic DMI
components are the results of a summation of pair interactions over all
neighbours. Although the Mn--Mn DMI have in general even larger magnitude than
the Fe--Fe interactions, their summation leads to a small total DMI due to
their oscillating behavior as a function of distance. This leads in the case of
MnGe to a significant compensation of all contributions. {For a more
realistic description of the experimental situation at finite temperature,
involving in particular non-collinear spin texture, Monte Carlo simulations
based on atomistic spin models might be important~\cite{PMB+14}.}

%%%%%%%%%%%%%%%%%%%%%%%%%%%%%%%%%%%%%%%%%%%%%%%%%%%%%%%%%%%%%%%%%%%%%%%%%%%%%%%
\begin{figure}[h]
\includegraphics[width=0.45\textwidth,angle=0,clip]{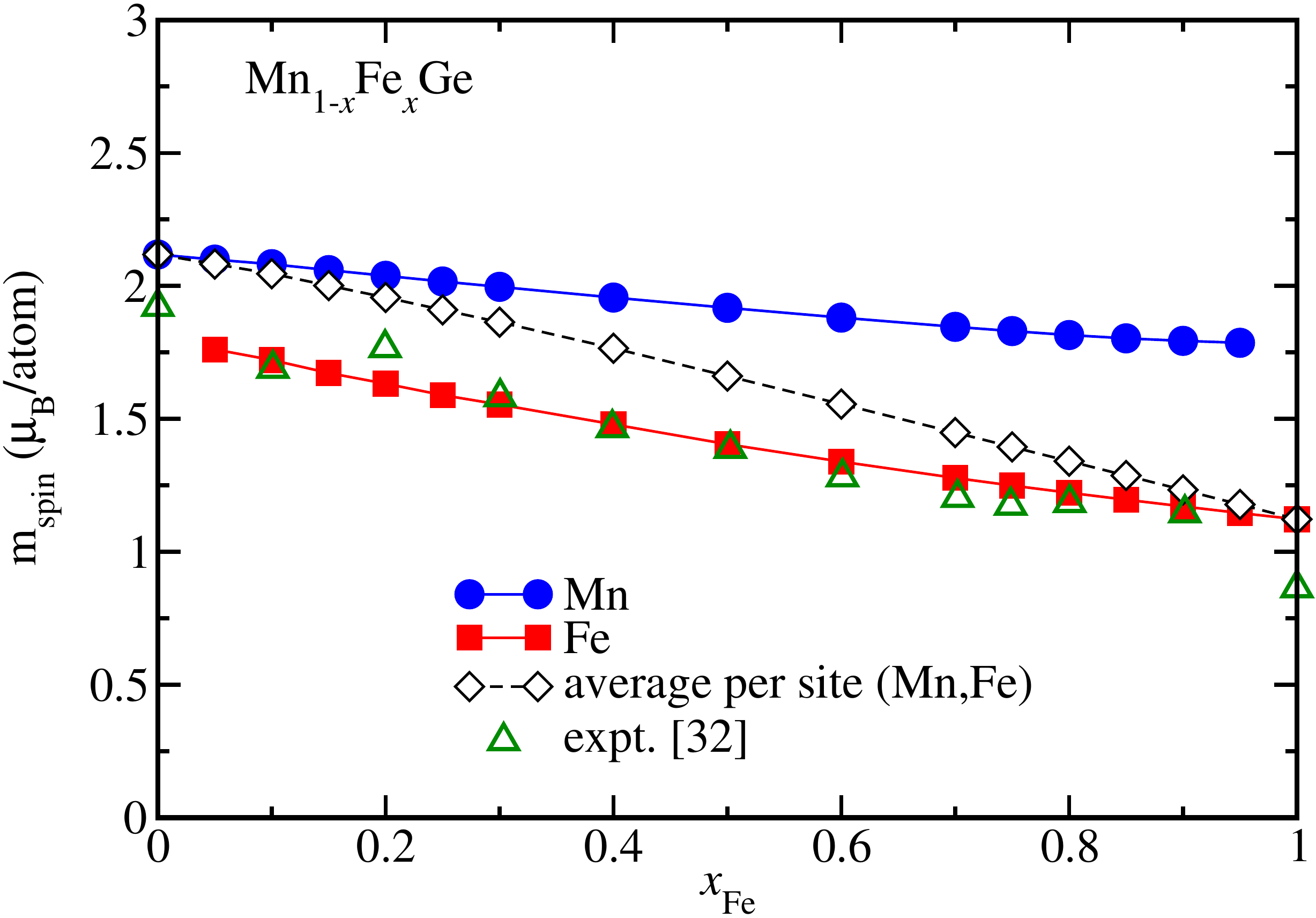}
\caption{\label{fig:M_spin} Spin magnetic moments of Mn (circles) and Fe
	(squares) in  Mn$_{1-x}$Fe$_x$Ge alloy, and the average magnetic moment
	per site (diamonds). Experimental results of Kanazawa \emph{et
al.} \cite{KST16} are shown as green triangles.}
\end{figure}
%%%%%%%%%%%%%%%%%%%%%%%%%%%%%%%%%%%%%%%%%%%%%%%%%%%%%%%%%%%%%%%%%%%%%%%%%%%%%%%

\subsection{Spin-orbit torque \label{ssec:SOT}}

\begin{figure}[h]
	\includegraphics[width=0.45\textwidth,angle=0,clip]{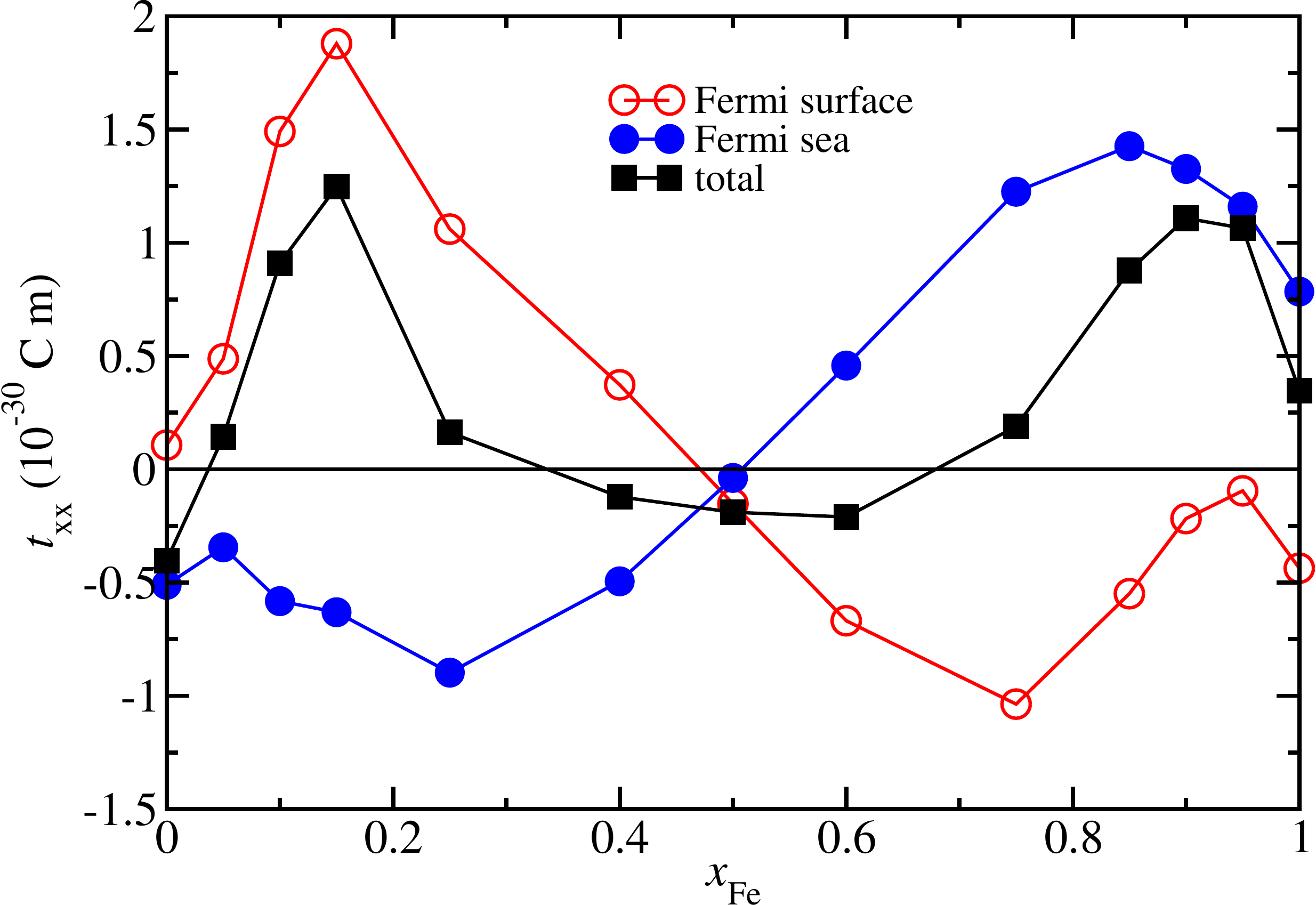}\;(a)
\includegraphics[width=0.45\textwidth,angle=0,clip]{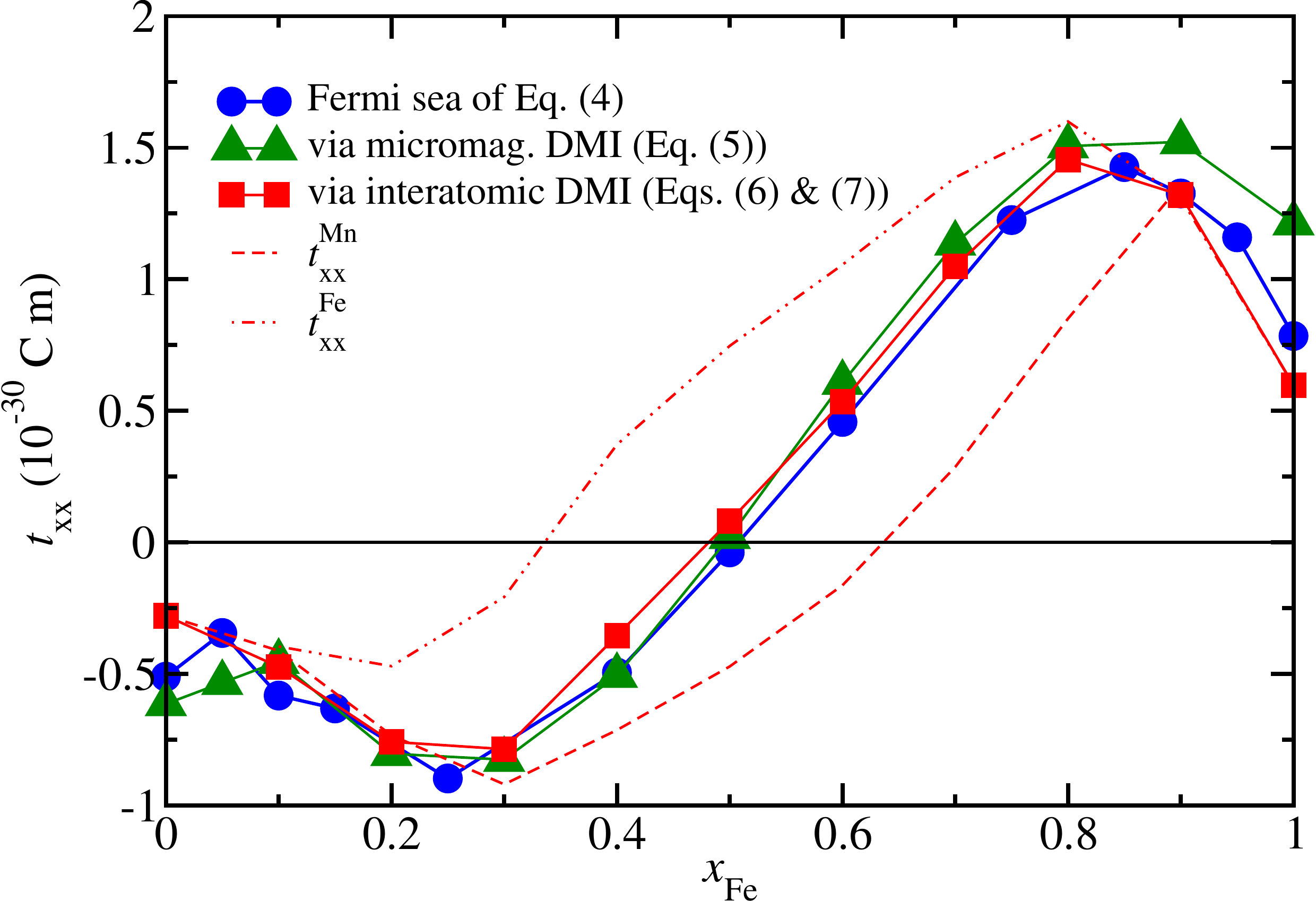}\;(b)
\caption{\label{fig:CMP_SOT} (a) Total torkance per unit cell (solid squares)
	as well as its Fermi surface (empty circles) and Fermi sea (filled
	circles) contributions in Mn$_{1-x}$Fe$_x$Ge calculated via the
	Kubo-Bastin formalism (\eqnref{Eq:Bastin-O1-O2-2}). (b) Comparison of
	the Fermi sea contribution to the torkance calculated via
	\eqnref{Eq:Bastin-O1-O2-2} (circles) with results obtained using the
	expressions \eqnref{Eq:txx_tau} (triangles) and
	\eqnref{Eq:SOT_micmag_vs_Heis} (squares).}
\end{figure}

The torkance tensor element $t_{\rm xx}(x)$ representing the spin-orbit torque
(SOT) calculated for Mn$_{1-x}$Fe$_x$Ge within the Kubo formalism \cite{WCS+16}
is represented in Fig.~\ref{fig:CMP_SOT} (a) by filled squares.  In contrast
to $D_{\rm xx}(x)$, it changes sign three times when $x$ increases. However, one
has to note that this behavior is caused by two contributions to the torkance,
showing a quite different concentration dependence: the Fermi surface
contribution from electronic states at the Fermi energy (open circles) and the
Fermi sea contribution due to all states below the Fermi energy (filled
circles).  Both contributions vary non-monotonously with $x$ and both change
sign at $x \sim 0.5$, having however an opposite slope in the vicinity of this
point. As a consequence, their combination leads to a partial cancellation in
the total torkance that has a completely different concentration dependence
when compared to the individual contributions.

\begin{figure}[h]
\includegraphics[width=0.45\textwidth,angle=0,clip]{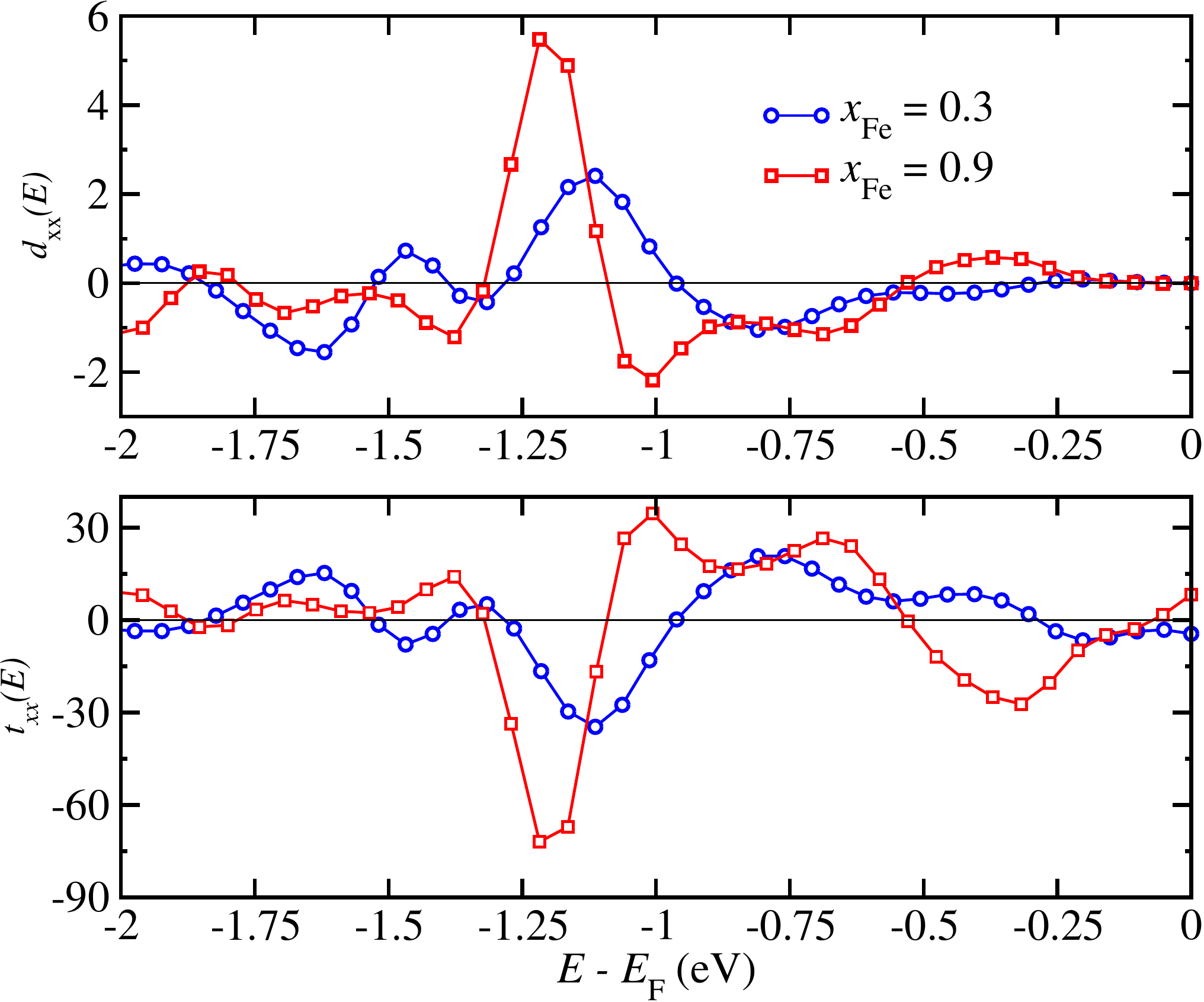}
\caption{\label{fig:DM_vs_E} Energy dependence of the DMI parameter $D_{\rm xx}(E)$
(upper panel) and the torkance $t_{\rm xx}(E)$ (lower panel) in
Mn$_{0.7}$Fe$_{0.3}$Ge (circles) and Mn$_{0.1}$Fe$_{0.9}$Ge (squares).
Integrated up to the Fermi energy $E_{\rm F}$, these functions give $t_{\rm xx}$
and $D_{\rm xx}$.
}  
\end{figure}

Despite similarities in the behavior of  $D_{\rm xx}(x)$ and the Fermi sea
torkance $t^{\rm sea}_{\rm xx}(x)$, they change sign at different $x$ values (0.8 and
0.5, respectively). To make a more detailed comparison, we calculate the Fermi
sea torkance using the expressions in Eqs.~(\ref{Eq:txx_tau}) and
(\ref{Eq:SOT_micmag_vs_Heis}). The results are plotted in
Fig.~\ref{fig:CMP_SOT} (b) (triangles and squares, respectively) in
comparison with the results based on the linear response expression
\eqnref{Eq:Bastin-O1-O2-2} (circles), demonstrating good agreement between all three
types of calculations. The difference in the concentrations when the
$D_{\rm xx}(x)$ and  $t^{\rm sea}_{\rm xx}(x)$ functions change sign can obviously be
attributed to the weighting factor $(E-\EF)$ in the expression for the DMI
\cite{ME17} which results in a different energy region for the dominating
contributions to the $D_{\rm xx}(x)$ function when compared to the torkance term
$t^{\rm sea}_{\rm xx}(x)$. This is demonstrated in Fig.~\ref{fig:DM_vs_E} that gives
the energy-resolved DMI parameter and the Fermi sea torkance for two different
Fe concentrations.  In addition, note that the contributions to
$t^{\rm sea}_{\rm xx}(x)$ associated with the alloy components Mn and Fe, shown in
Fig.~\ref{fig:CMP_SOT} (b) by dashed and dash-dotted curves, change
sign at different concentrations $x$.  Nevertheless, because of the strong
exchange interaction between these two components located on the same
sublattice, one has to discuss the component-averaged torkance when considering
the SOT in the alloy.

% %%%%%%%%%%%%%%%%%%%%%%%%%%%%%%%%%%%%%%%%%%%%%%%%%%%%%%%%%%%%%%%%%%%%%%%%%%%%%%%
% %%%%%%%%%%%%%%%%%%%%%%%%%%%%%%%%%%%%%%%%%%%%%%%%%%%%%%%%%%%%%%%%%%%%%%%%%%%%%%%

Finally, considering the Fermi surface and Fermi sea contributions to the SOT
separately in the pure limits, i.e., for the MnGe and FeGe compounds (see
Fig.~\ref{fig:CMP_SOT} (a)) one finds a different sign for these contributions.
This allows to conclude that the intrinsic torkance is mainly responsible for
the sign change of the SOT when the Fe concentration changes from 0 to 1.  It
is determined by the characteristics of the electronic structure discussed
above. On the other hand, in the case of disordered  Mn$_{1-x}$Fe$_x$Ge alloys
the extrinsic contributions to the SOT cannot be completely neglected. Although
small and only relevant at the Fermi surface, they are responsible together
with the intrinsic contribution for the concentration dependence of the SOT and
jointly determine the exact composition at which the torkance changes its sign.

%%%%%%%%%%%%%%%%%%%%%%%%%%%%%%%%%%%%%%%%%%%%%%%%%%%%%%%%%%%%%%%%%%%%%%%%%%%%%%%
%%%%%%%%%%%%%%%%%%%%%%%%%%%%%%%%%%%%%%%%%%%%%%%%%%%%%%%%%%%%%%%%%%%%%%%%%%%%%%%

\subsection{Anomalous and spin Hall conductivity \label{ssec:Hall}}

In order to have a more complete picture of the SOC-induced response to an
external electric field in Mn$_{1-x}$Fe$_x$Ge, we will briefly discuss
corresponding results for the transport properties anomalous Hall effect (AHE)
and spin-Hall effect (SHE) (see, e.g., Refs.~\cite{NSO+10} and \cite{SVW+15},
respectively). As the current-induced spin-orbit torkance, these phenomena are
caused by a SOC-induced spin asymmetry in the electron scattering. Because of
this one can expect certain correlations concerning their composition-dependent
behavior. 

% %%%%%%%%%%%%%%%%%%%%%%%%%%%%%%%%%%%%%%%%%%%%%%%%%%%%%%%%%%%%%%%%%%%%%%%%%%%%%%%
\begin{figure}[htb]
\includegraphics[width=0.45\textwidth,angle=0,clip]{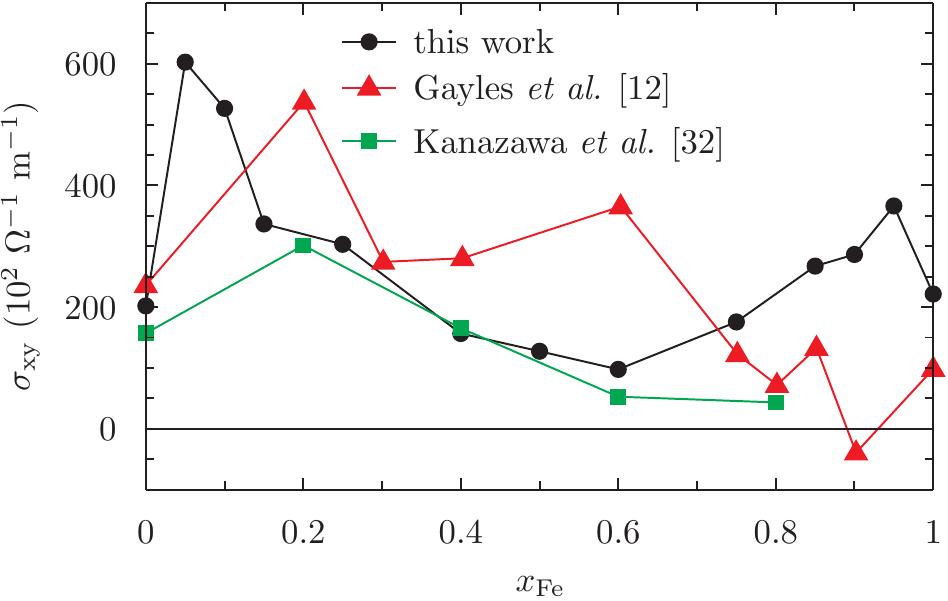}\;(a)
\includegraphics[width=0.45\textwidth,angle=0,clip]{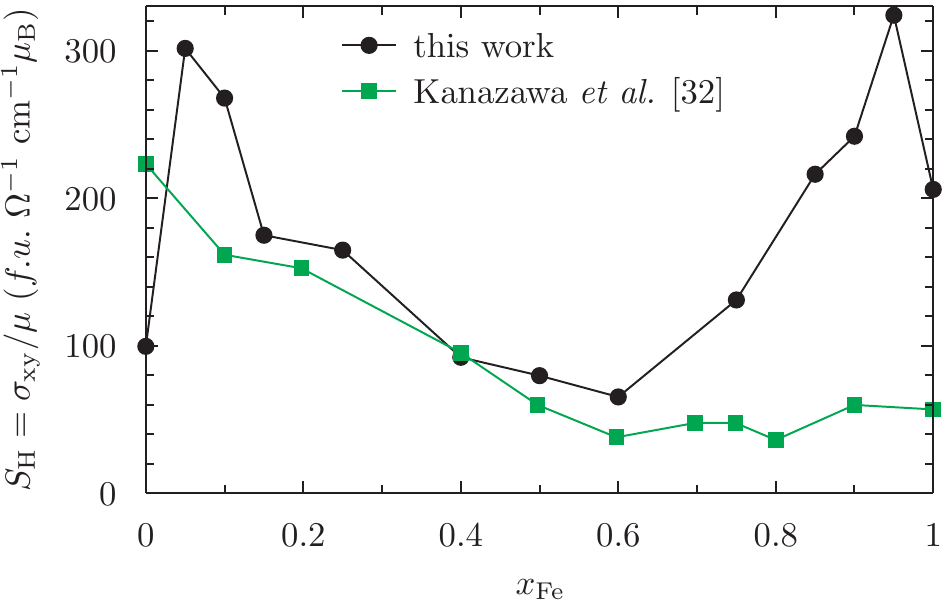}\;(b)
\caption{\label{fig:AHE} (a) Anomalous Hall conductivity calculated for
Mn$_{1-x}$Fe$_x$Ge via the CPA-Kubo-Bastin formalism (circles),
compared to calculations using the Berry curvature approach and the
Virtual Crystal Approximation \cite{GFS+15} (triangle) as well as to
low-temperature experimental data (squares) \cite{KST16}. (b) Anomalous
Hall coefficient calculated via the Kubo-Bastin equation (circles) compared to
experimental data at 50\,K (squares) \cite{KST16}.}
\end{figure}
% %%%%%%%%%%%%%%%%%%%%%%%%%%%%%%%%%%%%%%%%%%%%%%%%%%%%%%%%%%%%%%%%%%%%%%%%%%%%%%%

For the investigated alloy system Mn$_{1-x}$Fe$_x$Ge the anomalous Hall
conductivity (AHC) $\sigma_{\rm xy}$ calculated within the Kubo-Bastin
formalism (\eqnref{Eq:Bastin-O1-O2-2}) is given in Fig.~\ref{fig:AHE} (a) as
full circles. As can be seen, $\sigma_{\rm xy}$ does not change sign going from
MnGe to FeGe, in agreement with previous first-principles calculations
\cite{GFS+15} and experiment \cite{KST16}. Note that the chemical disorder is
treated on fundamentally different levels in the two theoretical approaches.
While the present work employs the Coherent Potential Approximation, the
results of Ref.~\onlinecite{GFS+15} are based on the Virtual Crystal
Approximation. This difference should be mainly responsible for the deviations
between the two sets of theory data visible in the upper panel of
Fig.~\ref{fig:AHE}, which are most pronounced on the Fe-rich side of the
concentration range where even the signs appears to differ. As will be shown
later, this is however not due to the extrinsic or incoherent contributions.
\tr{Unfortunately, reliable experimental data in this region could not be
obtained because both, the Hall as well as the longitudinal resistivity
are small under the experimental conditions \cite{KST16}.}

Comparison of the anomalous Hall coefficient $S_{\rm H} = \sigma_{\rm xy}/\mu$
to the experimental results of Kanazawa et al. \cite{KST16} in the lower panel
of Fig.~\ref{fig:AHE} shows good agreement for the Mn-rich side of the
concentration range (except for pure MnGe, see below), while deviations on the
Fe-rich side are quite large. Here one should note that the measurements were
performed at 50\,K while the calculations assume $T = 0\, K$, meaning in
particular perfect ferromagnetic order. As can be seen in Fig.~3 of
Ref.~\onlinecite{KST16} the temperature dependence of magnetization as well as
anomalous Hall conductivity is quite substantial for MnGe and even more so for
FeGe. \tr{As mentioned above for the anomalous Hall conductivity, the
experimental uncertainty is in addition rather high in the pure
Fe-limit. For a more detailed understanding of these discrepancies
investigations including the effects of finite temperature, sample
geometry, and non-collinear magnetic structure are necessary.}

\begin{figure}[t]
\includegraphics[width=0.45\textwidth,angle=0,clip]{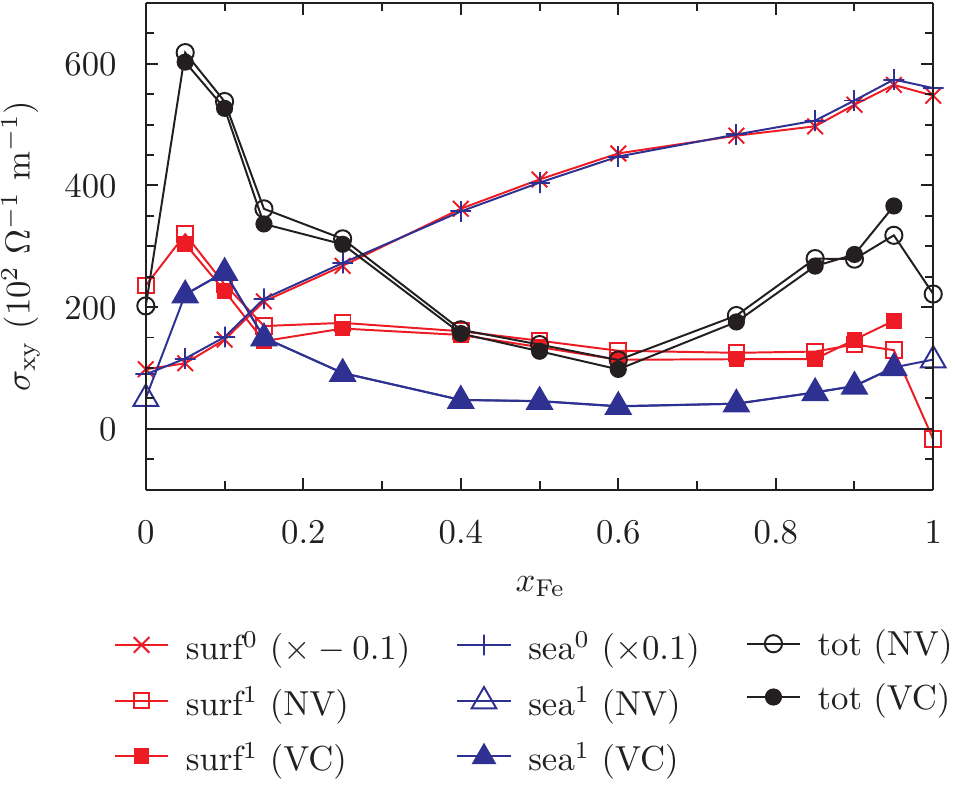}\;(a)\\
\vspace{0.2cm}
\includegraphics[width=0.45\textwidth,angle=0,clip]{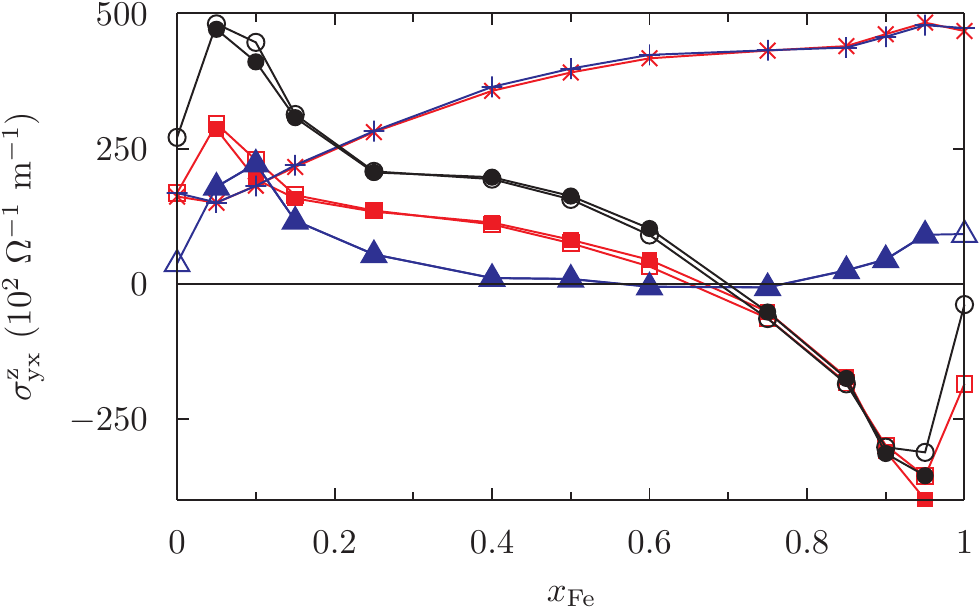}\;(b)
\caption{\label{fig:AHE_SHE} Anomalous (a) and spin (b) Hall
conductivities, $\sigma_{\rm xy}$ and $\sigma^z_{\rm xy}$, respectively, as functions of
$x$ in Mn$_{1-x}$Fe$_x$Ge calculated via the Kubo-Bastin formalism. Fermi
surface contributions are given in red, those from the Fermi sea in blue, and
their sum in black. The superscripts 0 and 1 indicate on- and off-site terms.
Results for the latter are shown ex- (NV) and including vertex corrections
(VC).}
\end{figure}

Having a closer look at the Kubo-Bastin equation, \eqnref{Eq:Bastin-O1-O2-2},
one can decompose the full response coefficient into several contributions with
distinct physical meaning. Most obviously, the two terms ${\cal R}^I_{\rm xy}$ and
${\cal R}^{II}_{\rm xy}$ differ in the absence or presence of contributions from
occupied states below the Fermi level, i.e., these are the Fermi surface and
Fermi sea terms, respectively. They are plotted in Fig.~\ref{fig:AHE_SHE}
(a) in red (Fermi surface) and blue (Fermi sea), further decomposed into
on-site (surf$^0$ and sea$^0$, crosses) and off-site (surf$^1$ and sea$^1$,
squares and triangles, respectively) contributions. For the latter results are
shown once excluding (NV, empty symbols) and once including the so-called
vertex corrections (VC, full symbols) arising from the difference in the
product of configuration-averaged Green functions versus the configuration
average of the product. These give rise to the so-called extrinsic or
incoherent contribution and are connected to the scattering-in term of the
Boltzmann equation \cite{But85}.

Comparing now the various terms one first of all notices that on-site terms are
large (note that they are scaled by a factor of (-)0.1), opposite in sign and
almost identical in magnitude, leading to an almost perfect cancellation.
Turning to the off-site terms one observes a similar concentration dependence
and a dominance of the Fermi surface contribution, except for $x \simeq 0.1$
and at the Fe-rich side of the concentration range. This means that the
anomalous Hall conductivity is dominated by the states at the Fermi level, in
particular for intermediate concentrations.  Obviously, already for this reason
a clear correlation between anomalous Hall coefficient and DMI strength, as
suggested by Kanazawa \emph{et al.} \cite{KST16}, is not supported by our findings.
Finally, the vertex corrections are, as observed before
\cite{TKD14,KCE15,WCS+16}, only relevant for the Fermi surface term and in this
system only noticeable in the dilute limits, particularly on the Fe-rich side.
Note that, as discussed before, there the density of states at the Fermi level
is largest and has predominantly $d$-like character.  Interestingly, the
seemingly diverging behavior for $x \rightarrow 0(1)$ is not caused by the
extrinsic contribution \cite{LGK+11}.
 
As the same spin-dependent scattering mechanisms are responsible for the SHE
and AHE, both, transverse spin and charge currents can be present in the FM
ordered Mn$_{1-x}$Fe$_x$Ge system. However, in contrast to $\sigma_{\rm xy}$
the transverse spin conductivity $\sigma^{\rm z}_{\rm xy}$ shown in
Fig.~\ref{fig:AHE_SHE} (bottom) does change its sign at $x \sim 0.7$. Thus, the
total transverse current should be dominated by opposite spin characters in
these limits. Interestingly, the AHC has a minimum of its absolute value close
to the Fe concentration corresponding to the sign change of the SHC. \tr{In
fact the Fermi sea contributions to $\sigma_{\rm xy}$ and $\sigma^{\rm z}_{\rm
xy}$ as well as both on-site terms behave very similarly over the entire
concentration range whereas the Fermi surface contributions agree only on the
Mn-rich side up to the minimum or sign change, respectively.}\\

% %%%%%%%%%%%%%%%%%%%%%%%%%%%%%%%%%%%%%%%%%%%%%%%%%%%%%%%%%%%%%%%%%%%%%%%%%%%%%%%
% %%%%%%%%%%%%%%%%%%%%%%%%%%%%%%%%%%%%%%%%%%%%%%%%%%%%%%%%%%%%%%%%%%%%%%%%%%%%%%%

The spin Hall conductivity of the Mn$_{1-x}$Fe$_x$Ge alloy system presented in
Fig.~\ref{fig:AHE_SHE} (bottom) as a function of Fe concentration changes sign
approximately at the same composition as the DMI parameter $D_{\rm xx}$ and,
accordingly, also the torkance $t_{\rm xx}$. However, one can again see a leading
role of the Fermi surface contribution to the spin-Hall conductivity, in
particular at the Fe-rich side after the sign change.  This implies that the
sign of the SHE conductivity is determined to a large extent by the character
of the states at the Fermi energy and their spin-orbit coupling, that changes
with concentration according to the discussion above. Note however, that in
pure FeGe the Fermi surface and Fermi sea contributions are of equal magnitude
but opposite sign, leading to their partial cancellation. Concerning the
importance of the vertex corrections the spin Hall conductivity behaves again
similar to the AHC, in as much as they are only present at the Fermi surface
and negligible over the entire concentration range considered here -- again
apart from the Fe-rich limit.
 
A more detailed analysis of the anomalous and spin Hall conductivities in terms
of underlying scattering mechanisms based on their scaling behavior w.r.t. to
the longitudinal (charge) conductivity in the dilute limits has been so far
precluded by the large numerical cost and is left for future work.  Note also,
that the anomalous and spin Hall conductivities in the present work were
calculated for the FM structure. Introducing a chiral non-collinear spin
texture, one can expect additional contributions from the topological anomalous
\cite{BDT04} and spin Hall \cite{YLB+15} effects, most likely displaying
different concentration-dependent features.\\

%\color{red}

\subsection{Symmetry considerations \label{ssec:Sym}}

	We conclude with a few remarks on magnetic symmetry and the
	corresponding shapes of the response tensors discussed above. The B20
	structure of the Mn$_{1-x}$Fe$_x$Ge alloy system has the (nonmagnetic)
	space group $P2_13$, for ferromagnetic order with magnetization along
	z (one of the $2_1$ axes) this leads to the magnetic space group (MSG)
	$P2_1^\prime2_1^\prime2_1$, the magnetic point group (MPG)
	$2^\prime2^\prime2$, and finally the magnetic Laue group (MLG)
	$m^{\prime}m^{\prime}m$ (or $2^\prime2^\prime2$ in the convention of
	Ref.~\onlinecite{Kle66}). The corresponding symmetry-allowed tensor
	forms for electrical ($\underline{\sigma}$) and spin
	($\underline{\sigma}^\xi$) conductivity~\cite{SKWE15}\footnote{Only
	given for polarization $\xi$ along z here, for x- and y-polarization see
        Ref.~\onlinecite{SKWE15}.} and the current-induced torkance~\cite{WCS+16} are
	\begin{equation}
        \label{eq:sigmas-MLGmpmpm}
	\underline{\sigma}^{(z)} = \left( \begin{array}{ccc} \sigma_{\rm xx}^{(z)} & \sigma_{\rm xy}^{(z)} & 0 \\
	                                                    -\sigma_{\rm xy}^{(z)} & \sigma_{\rm yy}^{(z)} & 0 \\
                                                                             0 & 0                 & \sigma_{\rm zz}^{(z)}
        \end{array} \right)
        \end{equation}
	and
	\begin{equation}
        \label{eq:torkance-MPG2p2p2}
	\underline{t} = \left( \begin{array}{ccc} t_{\rm xx} & t_{\rm xy} & 0 \\
	                                          t_{\rm yx} & t_{\rm yy} & 0 \\
	                                               0 & 0      & 0
        \end{array} \right) \, .
        \end{equation}
	Note, that this is not the highest symmetric FM-ordered structure as
	for $\vec m \parallel (111)$ (along the 3-fold axes) one would have MSG
	$R3$, MPG $3$ and MLG $\bar 3$, leading to the tensor shapes
	\begin{equation}
        \label{eq:sigmas-MLGbar3}
	\underline{\sigma}^{(z)} = \left( \begin{array}{ccc} \sigma_{\rm xx}^{(z)} & \sigma_{\rm xy}^{(z)} & 0 \\
	                                                    -\sigma_{\rm xy}^{(z)} & \sigma_{\rm xx}^{(z)} & 0 \\
                                                                             0 & 0                 & \sigma_{\rm zz}^{(z)}
        \end{array} \right)
        \end{equation}
	and
	\begin{equation}
        \label{eq:torkance-MPG3}
	\underline{t} = \left( \begin{array}{ccc} t_{\rm xx} & t_{\rm xy} & 0 \\
	                                         -t_{\rm xy} & t_{\rm xx} & 0 \\
	                                               0 & 0      & 0
        \end{array} \right) \, .
        \end{equation}
\begin{figure}[t]
\includegraphics[width=0.5\textwidth,angle=0,clip]{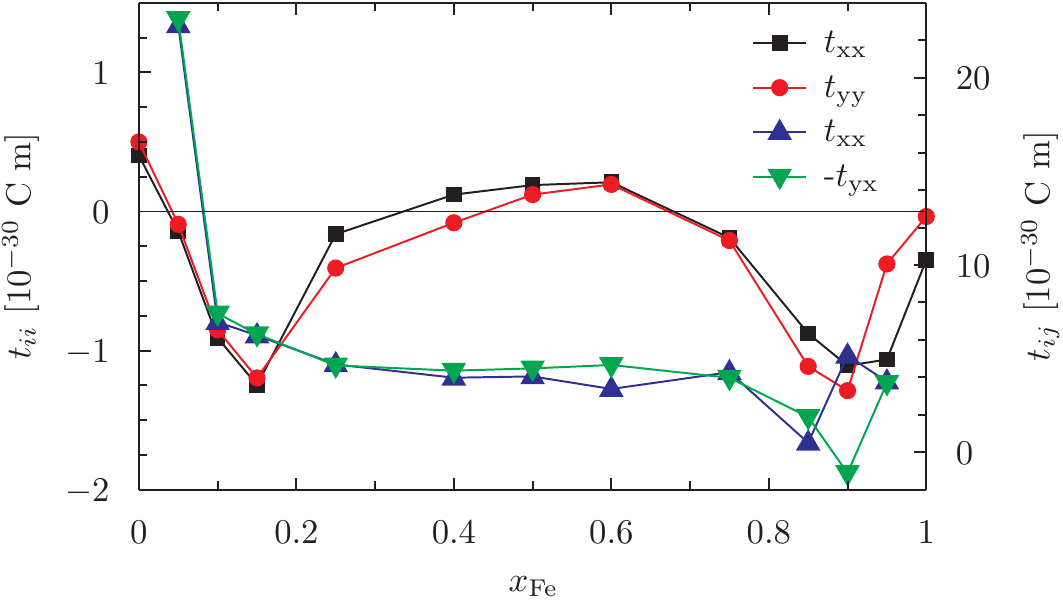}
\caption{\label{fig:tii_tij} Comparison of all non-zero torkance tensor elements as
	functions of $x$ in Mn$_{1-x}$Fe$_x$Ge calculated via the Kubo-Bastin
	formalism. The diagonal elements $t_{\rm xx}$ and $t_{\rm yy}$ (left y scale)
	are given as black squares and red circles, respectively, the
	off-diagonal torkances (right y scale) $t_{\rm xy}$ and $-t_{\rm yx}$ are given
        as blue up- and green down-facing triangles.}
\end{figure}
Fig.~\ref{fig:tii_tij} shows all non-zero tensor elements of $\underline{t}$
for $\vec m \parallel \rm z$ as chosen in this work. Apparently, the deviations
between the diagonal torkances $t_{\rm xx}$ and $t_{\rm yy}$ are negligibly
small over the whole concentration range, the largest differences occur once
more on the Fe-rich side. For the off-diagonal torkances $t_{\rm yx} \simeq
-t_{\rm xy}$ holds as well with the above exception. Note, that these torkances
in contrast to $t_{\rm xx}$ and $t_{\rm yy}$ only contain contributions from
the Fermi surface, as discussed before \cite{WCS+16}, and, as the diagonal
elements, are dominated by the intrinsic contribution.  Irrespective of the
magnetic point group ($m^{\prime}m^{\prime}m$ for $\vec m \parallel (001)$ or
$3$ for $\vec m \parallel (111)$), the diagonal elements are even, while the
off-diagonal ones are odd w.r.t. resversal of the magnetisation direction. The
same applies to both the electrical and the spin conductivity tensors.

%TODO:
%\begin{itemize}
%\item connection between spin conductivity ($\sigma_{\rm zx}^{\rm x}$) and even
%	torkance ($t_{\rm xx}$)?
%\item compare importance of contributions between SOT, AHC, and SHC:\\
%	on-site smaller than off-site for $t_{\rm xx}$, no cancellation of
%	$t^{0,surf}_{\rm xx}$ and $t^{0,sea}_{\rm xx}$; $t^{1,surf}_{\rm xx}$ and
%	$t^{1,sea}_{\rm xx}$ are equally important over whole concentration range.
%	In $t_{\rm xy}$ and $t_{\rm yx}$ the Fermi sea contributions are numerically
%	zero, the on-site contribution to the Fermi surface terms are
%	negligibly small.  large, opposite-signed $\sigma^{(z,)0,surf}_{\rm xy}$
%	and $\sigma^{(z,)0,sea}_{\rm xy}$, cancelling each other;
%	$\sigma^{(z,)1,surf}_{\rm xy}$ always dominates over
%	$\sigma^{(z,)1,sea}_{\rm xy}$, except for clean FeGe
%\end{itemize}

%\color{black}

\section{Summary \label{Sum}}

To summarize, we have presented results of calculations for the $D_{\rm xx}$
and $D_{\rm yy}$ components of the DMI vector in the B20 Mn$_{1-x}$Fe$_x$Ge
alloys as a function of Fe concentration. The sign change of this quantity
evidences the change of spin helicity at $x \approx 0.85$, in line with
experimental results as well as with theoretical results obtained by other
groups. Although the approach used in the present work is more appropriate for
disordered systems when compared to those used in the previous investigations,
all calculations demonstrate reasonable agreement, because of the
virtual-crystal-like behavior of the majority spin states \cite{GFS+15,KKAT16}.
In addition, we have discussed the concentration dependence of the total
spin-orbit torkance $t_{\rm xx}$ and its Fermi surface and Fermi sea
contributions.  It was shown that for all Fe concentrations both parts have the
same order of magnitude but their sign is opposite, leading to a significant
compensation.  By using different approaches to calculate the Fermi sea
contribution to the SOT its composition-dependent features in common with the
DMI were discussed. In the case of the AHE and SHE the calculated Fermi sea
contributions are rather small and the behavior of these effects as functions
of composition are determined mainly by the electronic states at the Fermi
level. The common SOC-induced mechanisms responsible for these effects, for the
investigated concentration range ($0.05 < x < 0.95$) these are predominantly of
intrinsic origin, result in the correlation of their dependence on the Fe
concentration. This is demonstrated by the finding that the minimum of the AHE
magnitude and the sign change of the SHC occur at approximately the same
composition.

\section{Appendix \label{App}}

According to the suggestion by Shilkova and Shirokovskii \cite{SST86,SS88},
the electron group velocity can be represented by the expression
%EEEEEEEEEEEEEEEEEEEEEEEEEEEEEEEEEEEEEEEEEEEEEEEEEEEEEEEEEEEEEEEEEEEEEEEEEE
\begin{eqnarray}
 \vec{v}_n(\vec{k})  &=& \frac{\partial
     \lambda_n(E,\vec{k})}{\partial \vec{k}}}\bigg|_{E = E_n(\vec{k})}\bigg/{\frac{\partial
     \lambda_n(E,\vec{k})}{\partial E} \bigg|_{E = E_n(\vec{k})}  \;. 
\label{Eq:velocity} 
\end{eqnarray}
%EEEEEEEEEEEEEEEEEEEEEEEEEEEEEEEEEEEEEEEEEEEEEEEEEEEEEEEEEEEEEEEEEEEEEEEEEE
Here $\lambda_n(E,\vec{k})$ are the eigenvalues of the KKR matrix
$\underline{M}(E,\vec{k}) = \underline{\tau}^{-1}(E,\vec{k})$  that are
determined by solving the eigenvalue problem \cite{GFP+11} 
%EEEEEEEEEEEEEEEEEEEEEEEEEEEEEEEEEEEEEEEEEEEEEEEEEEEEEEEEEEEEEEEEEEEEEEEEEE
\begin{eqnarray}
\underline{M}(E,\vec{k}) \vec{b}^{n}(E,\vec{k}) &=& \lambda_n
\vec{b}^{n}(E,\vec{k}) \\ \nonumber 
\label{Eq:eigenvalue1} 
\end{eqnarray}
%EEEEEEEEEEEEEEEEEEEEEEEEEEEEEEEEEEEEEEEEEEEEEEEEEEEEEEEEEEEEEEEEEEEEEEEEEE
and vanish at  $E = E_n(\vec{k})$ corresponding to  zeros of the determinant
$||\underline{M}(E,\vec{k})||$.  Here $b^{n}_{\Lambda}(\vec{k})$ are the
associated eigenvectors.  With this one arrives at the expression 
%EEEEEEEEEEEEEEEEEEEEEEEEEEEEEEEEEEEEEEEEEEEEEEEEEEEEEEEEEEEEEEEEEEEEEEEEEE
\begin{eqnarray}
&& \vec{b}^{n\dagger}(\vec{k}) \frac{\partial
  \underline{\tau}(E,\vec{k}) }{\partial \vec{k}} \vec{b}^{n}(\vec{k})|_{E =
  E_n(\vec{k})} 
\nonumber \\
 &=&
\vec{v}_n(\vec{k}) \vec{b}^{n\dagger}(\vec{k}) \frac{\partial
 \underline{\tau}(E,\vec{k})}{\partial E} \vec{b}^{n}(\vec{k})|_{E = E_n(\vec{k})} \;.
\label{Eq:eigenvalue2} 
\end{eqnarray}
%EEEEEEEEEEEEEEEEEEEEEEEEEEEEEEEEEEEEEEEEEEEEEEEEEEEEEEEEEEEEEEEEEEEEEEEEEE
Finally, use is made of the relation for the group velocity 
\cite{SST86,SS88} 
%EEEEEEEEEEEEEEEEEEEEEEEEEEEEEEEEEEEEEEEEEEEEEEEEEEEEEEEEEEEEEEEEEEEEEEEEEE
\begin{eqnarray}
 \vec{v}_n(\vec{k})  &=& \sum_{\Lambda,\Lambda'}b^{n\dagger}_{\Lambda}(\vec{k})
(c \boldsymbol{\alpha}_{\Lambda,\Lambda'}) b^{n}_{\Lambda'}(\vec{k}) 
\label{Eq:velocity2} 
\end{eqnarray}
%EEEEEEEEEEEEEEEEEEEEEEEEEEEEEEEEEEEEEEEEEEEEEEEEEEEEEEEEEEEEEEEEEEEEEEEEEE
with
%EEEEEEEEEEEEEEEEEEEEEEEEEEEEEEEEEEEEEEEEEEEEEEEEEEEEEEEEEEEEEEEEEEEEEEEEEE
\begin{eqnarray}
 \boldsymbol\alpha_{\Lambda,\Lambda'} & = & \int_{\Omega} d^3r  \,
Z^{\times}_{\Lambda}(\vec{r},E) \,  \boldsymbol{\alpha}  Z_{\Lambda'}(\vec{r},E)  \label{Eq:ME1}
 \label{Eq:MEalph}
\end{eqnarray}
%EEEEEEEEEEEEEEEEEEEEEEEEEEEEEEEEEEEEEEEEEEEEEEEEEEEEEEEEEEEEEEEEEEEEEEEEEE
where $c$ is the speed of light and $\boldsymbol{\alpha}$ is the vector of
Dirac matrices, that represents the relativistic current operator
$\hat{\vec{j}} = -e\hat{\vec{v}} = -ec\boldsymbol{\alpha}$. With this one
finally arrives at the relationship between \eqnref{Eq:txx_tau} and the Fermi
sea term in \eqnref{Eq:Bastin-O1-O2}.

\begin{acknowledgments}
Financial support by the DFG via SFB 689 (Spinph\"anomene in reduzierten
Dimensionen) and SFB 1277 (Emergent Relativistic Effects in Condensed Matter -
From Fundamental Aspects to Electronic Functionality) is gratefully
acknowledged.
\end{acknowledgments}

%\bibliography{FeMn_Ge,SKYRMION,NEW_DMI,/opt/ak/bib/akhelit,SPRKKR7}

\end{document}